\documentclass[12pt,preprint]{aastex}
\newcommand{\mdot}{M$_{\odot}$ yr$^{-1}$}
\newcommand{\ldot}{L$_{\odot}$}
\newcommand{ \um}{$\mu$m~}
\newcommand{ \ums}{$\mu$m}
\def\kmsMpc{\ifmmode {\rm\,km\,s^{-1}\,Mpc^{-1}}\else
    ${\rm\,km\,s^{-1}\,Mpc^{-1}}$\fi}
\shorttitle{ Infrared Emission Lines for Starbursts and AGN}
\shortauthors{Sargsyan et al.}

\begin{document}

\title{Star Formation Rates from [CII] 158 \um and Mid Infrared Emission Lines for Starbursts and AGN\footnote{Based on observations with the $Herschel$ Space Observatory, which is an ESA space observatory with science instruments provided by European-led Principal Investigator consortia and with important participation
from NASA.}} 

\author{ L. Sargsyan\altaffilmark{1}, A. Samsonyan\altaffilmark{2}, V. Lebouteiller\altaffilmark{3,1}, D. Weedman\altaffilmark{1}, D. Barry\altaffilmark{1}, J. Bernard-Salas\altaffilmark{5}, J. Houck\altaffilmark{4}, and H. Spoon\altaffilmark{1}}

\altaffiltext {1}{Center for Radiophysics and Space Research, Cornell University, Ithaca, NY 14853; sargsyan@isc.astro.cornell.edu, dweedman@isc.astro.cornell.edu }
\altaffiltext{2}{Byurakan Astrophysical Observatory, Byurakan, Armenia}
\altaffiltext {3} {Laboratoire AIM, CEA/DSM-CNRS-Universite Paris Diderot, DAPNIA/Service d'Astrophysique, Saclay, France}
\altaffiltext {4}{Astronomy Department, Cornell University, Ithaca,
NY 14853}
\altaffiltext{5} {Open University, Department of Physical Sciences, Milton
Keynes, MK7 6AA, UK
}
%\altaffiltext{4} {Hamburger Sternwarte, Hamburg, Germany}
%\altaffiltext{5} {Universite Paris XI, Orsay, France}

\begin{abstract}

A summary is presented for 130 galaxies observed with the $Herschel$ PACS instrument to measure fluxes for the [CII] 158 \um emission line.  Sources cover a wide range of active galactic nucleus to starburst classifications, as derived from polycyclic aromatic hydrocarbon (PAH) strength measured with the $Spitzer$ Infrared Spectrograph.  Redshifts from [CII] and line to continuum strengths (equivalent width of [CII]) are given for the full sample, which includes 18 new [CII] flux measures.  Calibration of $L([CII)])$ as a star formation rate (SFR) indicator is determined by comparing [CII] luminosities with mid-infrared [NeII] and [NeIII] emission line luminosities; this gives the same result as determining SFR using bolometric luminosities of reradiating dust from starbursts:  log SFR = log $L([CII)])$ - 7.0, for SFR in \mdot~ and $L([CII])$ in \ldot.  We conclude that $L([CII])$ can be used to measure SFR in any source to a precision of $\sim$ 50\%, even if total source luminosities are dominated by an AGN component. The line to continuum ratio at 158 \ums, EW([CII]), is not significantly greater for starbursts (median EW([CII]) = 1.0 \ums) compared to composites and AGN (median EW([CII]) = 0.7 \ums), showing that the far infrared continuum at 158 \um  scales with [CII] regardless of classification. This indicates that the continuum at 158 \um also arises primarily from the starburst component within any source, giving log SFR = log $\nu L_{\nu}$(158 \ums) - 42.8 for SFR in \mdot~ and $\nu L_{\nu}$(158 \ums) in erg s$^{-1}$.

\end{abstract}


\section{Introduction}

To discover and understand dusty galaxies at the highest redshifts, the emission of [CII] 158 \micron~is the single most important atomic line feature because it is the strongest far-infrared line in most sources \citep{sta91, mal97, nik98, luh03, bra08, far13} and also provides the best opportunity for high redshift determinations and source diagnostics using submillimeter and millimeter spectroscopic observations.  Already, for example, [CII] has been measured with the Atacama Large Millimeter Array (ALMA) in sources with 4 $<$ z $\sim$ 6 \citep{swi12,wan13,car13}, demonstrating that large scale star formation extends to high redshifts.  The [CII] line should be a diagnostic of star formation because it is primarily associated with the photodissociation region (PDR) surrounding starbursts \citep{tie85,hel01,mal01,mei07}.

The objective of our studies is to compare characteristics of [CII] emission with classifications and luminosities of sources determined from mid-infrared spectra, so that [CII] can be used alone as a quantitative diagnostic when no other spectroscopy is available.  This is especially important for dusty sources at high redshifts, when the rest frame far-infrared can be observed in submillimeter and millimeter wavelengths, but other rest frame wavelengths are not observationally accessible.

Our observations consist of a sample of 130 sources which include a wide range of active galactic nucleus (AGN) through starburst (SB) classifications, using [CII] observations made with the Photodetector Array Camera and Spectrometer (PACS; Poglitsch et al. 2010) on the $Herschel$ Space Observatory \citep{pil10}.  The [CII] results are compared to mid-infrared spectroscopic diagnostics determined with the $Spitzer$ Infrared Spectrograph (IRS; Houck et al. 2004).  The sample is chosen to minimize effects of differing spatial resolutions when comparing $Herschel$ far infrared observations with $Spitzer$ mid-infrared slit spectroscopy by choosing sources that are unresolved.  Sample selection, classification, and results for [CII] fluxes for 112 sources are described in \citet{sar12}; hereinafter ``Paper 1".  

In that paper, it was found that the [CII] line flux correlates closely with the flux of the polycyclic aromatic hydrocarbon (PAH) 11.3 \um feature independently of AGN/starburst classification, log [f([CII] 158 \ums)/f(11.3 \um PAH)] = -0.22 $\pm$ 0.25.  As a result, we concluded that [CII] line flux measures the same photodissociation regions (PDRs) associated with starbursts as the PAH feature \citep{pee04} and, therefore, measures the starburst component of any source.  A calibration of star formation rate (SFR) was derived by comparing [CII] luminosity $L([CII])$ to total infrared luminosity $L_{ir}$ using the precepts of \citet{ken98}, with the result that log SFR = log $L([CII)])$ - 7.08 $\pm$ 0.3, for SFR in \mdot~ and $L([CII])$ in \ldot.

A comprehensive summary is given by De Looze et al.(2014) of all available [CII] measurements (including those in Paper 1) and their use as a SFR indicator through calibration with continuum luminosities (including both ultraviolet and infrared).  Their emphasis is on determining uncertainties within the SFR calibration arising from cosmic variance and on discussing why differences can arise among sources.  For starbursts, the summary results are very similar to ours in Paper 1, with the same calibration and similar scatter for the relation between $L([CII])$ and SFR. 

The uncertainties arising from cosmic variance among sources do not include systematic uncertainties in the assumptions used to transform total radiated luminosities from a star forming region into SFRs.  It is important, therefore, to determine independent calibrations of $L([CII)])$ compared to SFR which depend on different precepts.  The primary objective of the present paper is to obtain an independent calibration of $L([CII)])$ as a SFR indicator by comparing [CII] luminosities with mid-infrared [NeII] and [NeIII] emission line luminosities, calibrated for SFR by \citet{ho07}, using new measurements of high resolution IRS spectra (section 3.2).  In addition, we determine the [CII] line to continuum ratio at 158 \um and use this to calibrate SFR from the continuum luminosity at 158 \um (section 3.3); this allows an independent comparison to a previous calibration of SFR from 160 \um luminosity \citep{cal10}.  As part of these analyses, we also provide new [CII] observations of an additional 18 sources and provide new redshifts for all sources determined using only the [CII] emission line.  

\section{Observations}

\subsection{Selection and Classification of Sources}

As summarized in Paper 1, the sample selected for [CII] observations with $Herschel$ PACS derives from sources having both $Spitzer$ IRS spectra and full photometry with the Infrared Astronomical Satellite (IRAS), and for which f$_{\nu}$(IRS)/f$_{\nu}$(IRAS) comparisons at 25 \um indicate that sources are unresolved (because comparable source fluxes are measured with the slit of the IRS compared to the large beam of IRAS).  Sources encompass a wide range of classification, from AGN through composite to starburst, derived uniformly from the strength of the PAH 6.2 \um feature in IRS spectra. 

Our AGN/starburst classification, based on strength of the 6.2 \um PAH feature, is described in Paper 1 and in \citet{sar11}.  The classification criterion is similar to that used in many previous studies, although different authors use different PAH features and different methods for measuring strength \citep[e.g.][]{gen98,lau00,des07,vei09,tom10,wu10,sti13}.  The primary motive for using the 6.2 \um feature instead of the stronger 11.3 \um feature is to compare with sources having sufficiently high redshifts that the 11.3 \um feature is not visible in IRS spectra.  Our measurement is simply the observed EW(PAH 6.2 \ums) determined in a uniform fashion without any modeling assumptions regarding the mix of PAH features and/or the true level of underlying dust continuum.  This measurement is made with the SMART software for IRS spectra \citep{hig04}, and the EW(PAH 6.2 \ums) is determined using a Gaussian fit to the PAH feature and a linear fit to the continuum beneath the feature within the range 5.5 \um to 6.9 \ums.   Measurements from IRS low resolution spectra and the empirical correlation of EW(PAH 6.2 \ums) with source classification are given in \citet{sar11}.  All spectra are available in the CASSIS spectral atlas \citep{leb11}\footnote{http://cassis.astro.cornell.edu. The Cornell Atlas of Spitzer IRS Spectra (CASSIS) is a product of the Infrared Science Center at Cornell University.}.

\begin{figure}
\figurenum{1}
\includegraphics[scale=0.9]{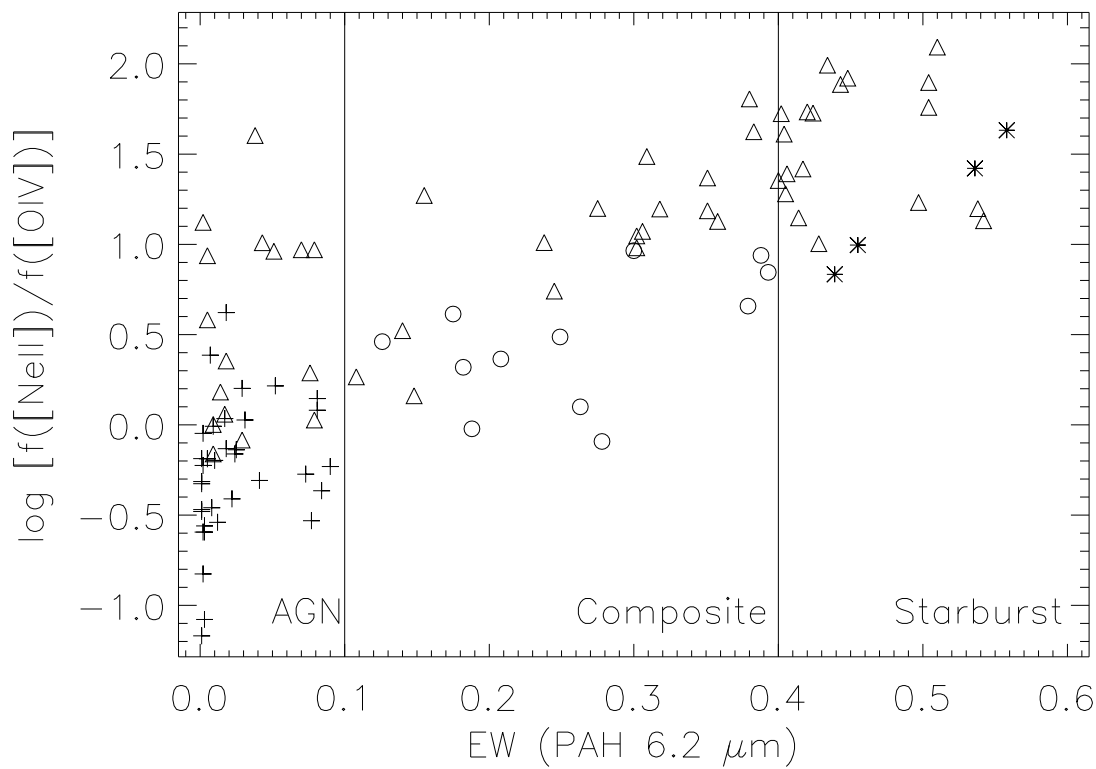}
\caption{ Ratio of [NeII] 12.81 \um to [OIV] 25.89 \um emission line fluxes compared to equivalent width of PAH 6.2 \um feature in \ums.  Dividing lines show the division into classifications based on EW(PAH 6.2 \ums) used throughout this paper.  Crosses are sources classified as pure AGN, asterisks as pure starbursts, and circles are composites with contribution from both.  Triangles are lower limits, in which the [OIV] feature is not measured. } 

\end{figure}

%(Genzel, Laurent, Veilleux, Desai, Tommasin, EVANS, stierwalt? FOR GOALS DIAGNOSITICS??  (Laurent,Evans,Stierwalt)  are    (Sargsyan et al.) VIELLEUX 2009 FOR EMISSION LINE AND PAH 6.2   RUPKE 2009 FOR DIAGNOSTICS IN THEIR QUEST SURVEY.  VERY SIMILAR TO THE ONE PLOT WE SHOW HERE. ALSO CHECK TOMMASIN 2010,2012 FOR HI RES, and mention Veilluax, Farrah

For purposes of the present paper, the most important use of these diagnostics is to select sources confidently classified as pure starbursts so these can be used to determine various relations among [CII], infrared luminosity, PAH luminosities, and emission line luminosities that apply to pure starburst sources.  These relations provide the calibration of SFR.  The composite and AGN sources then illustrate changes that occur with increasing fractions of AGN luminosity added to the starburst. 

The classification utilized in the following discussion is summarized in Figure 1 using sources discussed below having both PAH and emission line measurements.  The quantitative division among ``pure AGN" and ``pure starbursts", with the intermediate ``composite" systems is illustrated.  The EW boundaries which are shown derive by combining optical classifications and mid-infrared template spectra to determine classification boundaries as summarized in \citet{sar11} and in Paper 1.  Starbursts have EW(PAH 6.2 \ums) $>$ 0.4 \ums, composites have 0.1 \um $<$ EW(PAH 6.2 \ums) $<$ 0.4 \um, and AGN have EW(PAH 6.2 \ums) $<$ 0.1 \ums.   Pure starbursts are defined as systems with $>$ 90\% of the mid-infrared luminosity arising from a starburst, and pure AGN as systems with $>$ 90\% of the mid-infrared luminosity arising from the AGN (Paper 1).  Composites have mixtures of these components.  For the full sample of 130 sources in this paper, these criteria give 60 AGN, 36 composites, and 34 starbursts.

The summary of results in Figure 1 comparing the ratio of [NeII] 12.81 \um to [OIV] 25.89 \um emission line fluxes with the classification derived from PAH EW illustrates the conclusions from many previous studies \citep[e.g.][]{voi92,gen98,stu02,ver03,far07} that PAH strength correlates with level of ionization as seen in emission lines, which confirms the AGN/starburst classification derived from PAH strength.   Both the EW(PAH 6.2 \ums) from the starburst PDR and the relative [NeII] emission line strength from the starburst HII region decrease as the starburst component decreases, because the AGN adds additional hot dust continuum in the mid-infrared that diminishes EW(PAH 6.2 \ums) and also adds increasing strength of the higher ionization [OIV] (the ionization potential required to produce NeII is 21.6 eV compared to 55 eV to produce OIV).

\subsection{[CII] Observations and Uncertainties}

The [CII] observations are described in Paper 1, where examples of observed line profiles are illustrated.  All [CII] observations were made using the $Herschel$ PACS instrument \citep{pog10} for line spectroscopy in point source chop nod mode with medium throw.  [CII] line fluxes are obtained by summing fluxes in the nine central equivalent spatial pixels, or ``spaxels", produced by the PACS image slicer\footnote{HERSCHEL-HSC-DOC-0832, Version 2; http://herschel.esac.esa.int/Docs/PACS/pdf/pacs-om.pdf}.  A correction communicated to us by the PACS calibration team is applied for flux that would fall outside these spaxels for an unresolved source, correcting the 9 spaxel flux by a factor of 1.16($\lambda$/158 \ums)$^{0.17}$, with wavelength $\lambda$ depending on redshift.  Data reduction in Paper 1 was done with version 8 of the $Herschel$ Interactive Processing Environment (HIPE), and the ``PACSman" software \citep{leb12} is used for fitting line profiles and continuum.  

An additional 18 sources were subsequently observed in program lsargsya-OT2.  These [CII] fluxes were derived in a similar way and are given in Table 1. Results for the total sample of 130 sources are summarized in Table 2 where [CII] fluxes for the 112 sources from Paper 1 are also reproduced.  Line fluxes are determined from the flux within a fitted Gaussian profile except for 12 sources (noted in Table 2) for which the [CII] profile is asymmetric or shows component structure.  For these 12, the total line flux is the integrated flux including all components rather than the flux within a single Gaussian fit.

All results are determined with HIPE version 8. The new sources were also extracted with HIPE version 10, which has different calibration from version 8.   By reducing new sources with both v8 and v10, we find a systematic flux difference such that v8 fluxes are a factor of 1.07 brighter compared to v10.  This difference is noted in Table 1, but fluxes listed for the new sources arise from v8 to be consistent with results in Paper 1. 

PACS observations of the [CII] line are made in a continuous scanning mode that produces a "data cloud" containing a large number of separate flux measurements within individual wavelength elements because the spectrometer is read out every 1/8 s (Poglitsch et al. 2010 and PACS Observer's Manual) during the total integration time of 579 s for these observations.  The PACSman uncertainties at each wavelength are determined from the dispersion among the individual measurements within the data cloud.  The S/N of the overall profile fit is determined using the uncertainties within these individual wavelength elements.  Examples of profiles and fits are shown in Paper 1, including a noisy profile with plotted uncertainties. 

Uncertainties of individual [CII] emission line fluxes are determined by the signal to noise (S/N) of the profile fit by PACSman in the brightest spaxel, with uncertainties listed in Paper 1 and Table 1 individually by source.  S/N is defined as the (line flux)/(uncertainty in line flux) arising from the profile fitting. If S/N $<$ 3, fluxes were listed as upper limits in Paper 1 and are shown as upper limits in plots within the present paper.  No new sources in Table 1 are limits.  Median S/N for the remaining sources indicates a median line flux uncertainty of $\pm$ 15\%.  The systematic uncertainty for [CII] fluxes depends on PACS flux calibration, estimated as $\pm$ 12\% in the PACS Spectroscopy performance and calibration document PICC-KL-TN-041.  (The offset we find between HIPE v8 and v10 reductions is within this uncertainty.)  Combining these two sources of uncertainty yields a typical combined uncertainty of $\pm$ 20\% for [CII] fluxes.  

In paper 1, redshifts were listed based on previously determined optical redshifts, verified by IRS mid-infrared emission lines.  In Tables 1 and 2 of the present paper, we list new redshifts derived from the [CII] line only.  Redshifts generally agree to $\la$ 0.002, but the [CII] redshifts are given here  to allow further study of any systematic differences between [CII] emitting regions and optical or mid-infrared regions.  

Uncertainties in measured [CII] line velocities are dominated by centering effects, as described in the PACS Observer's Manual.  Depending on precisely where within a spaxel the source is centered, the observed wavelength of the emission line can change by $\sim$ $\pm$ 40 km s$^{-1}$.  To estimate empirically the uncertainties arising both from this effect and from fitting the line profile, we compare in Figure 2 the [CII] centroid velocity within the brightest spaxel relative to the velocity within the second brightest spaxel (only for sources in which both spaxels have S/N $>$ 3). This plot does not include the 12 sources noted in Table 2 with asymmetric or component structure to the [CII] line profile, because these sources may really have spatially different components on a sub-spaxel scale.  

From Figure 2, the brightest spaxel systematically gives the same velocity as other spaxels; the median velocity offset between the brightest and second brightest is less than 1 km s$^{-1}$.  The random errors from source position and profile fitting are determined as the one sigma dispersion among the velocity differences shown between the brightest spaxel and the second brightest, which gives a dispersion of $\pm$ 50 km s$^{-1}$.  This is our measurement of empirical uncertainty for the measured [CII] velocities.  This dispersion is much larger than the formal centroiding uncertainty of the line fit for the brightest spaxel, typically only a few km s$^{-1}$ in the PACSman fits, indicating that velocity uncertainties are dominated by positional effects of the source within the spaxel. The final redshift measurement which is reported in Tables 1 and 2 is that of only the brightest spaxel.  

In section 3.3, the line to continuum ratio (equivalent width, EW) for the [CII] line is discussed.  This is important for understanding the relation between far infrared dust continuum reradiation and the origin of the [CII] emission line.  Uncertainties in EW arise from both the [CII] emission as uncertainties in the line flux, and as uncertainties in the level of the underlying continuum.  Usually, the latter is a much larger source of uncertainty because source continua are faint compared to the background.  These two uncertainties as determined by PACSman are summed quadratically to give total uncertainty in the EW which is listed in Table 2. 

Comparisons of [CII] with PAH fluxes in Paper 1 and below are made using the 11.3 \um PAH feature because this is the strongest PAH feature in IRS spectra.  The flux ratios for the new sources are given in Table 1, using measurements of the 11.3 \um feature taken from \citet{sar11} as determined with a Gaussian fit to the PAH feature and a linear fit to the continuum beneath the feature from 10.5 \um to 12 \ums.

\begin{figure}
\figurenum{2}
\includegraphics[scale=0.9]{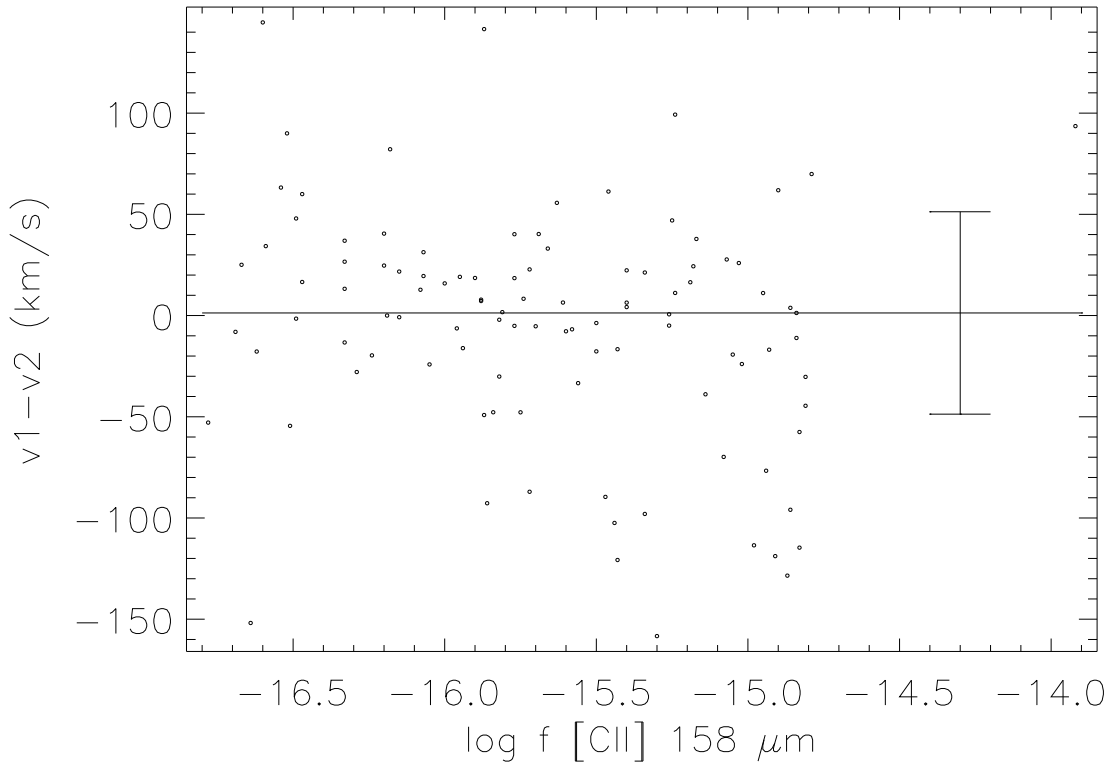}
\caption{ Difference in measured [CII] line profile velocities in km s$^{-1}$ between the profile centroid velocity in the brightest spaxel (v1) compared to the velocity in the second brightest spaxel (v2), using only sources in which both spaxels have S/N $>$ 3 for the measured line flux; line flux is shown in units of W m$^{-2}$.  The empirical uncertainty of final [CII] velocity measurements is taken as the one sigma dispersion among these velocity differences of $\pm$ 50 km s$^{-1}$, which includes uncertainties arising both from profile fitting and from location of the source within a spaxel. This plot does not include the 12 sources with asymmetric or component structure for which we only use the velocities from the brightest spaxel, because these are sources which may have real spatially different components. } 

\end{figure}

\subsection{IRS Emission Line Observations and Uncertainties}

In paper I, [CII] observations were compared with PAH strength using IRS low resolution spectra.  All objects in that paper also have IRS high resolution spectra, which derive from archival observations in various $Spitzer$ observing programs \citep[e.g.][]{far07,vei09,tom10,pet11}.  For many sources, line fluxes have been previously published.  For the present paper, we remeasure emission lines in these high resolution spectra with the goal of a uniformly measured data set that uses the final IRS flux calibrations and noise masks; the new measurements also provide an empirical determination of line flux uncertainties that arise from the profile fitting process and noise removal.  

Line fluxes are measured using Gaussian fits in the SMART analysis program \citep{hig04}, beginning with the post-Basic Calibrated Data products in the $Spitzer$ Heritage Archive\footnote{http://irsa.ipac.caltech.edu/data/SPITZER/docs/spitzerdataarchives/}.  The line fit is on top of the underlying continuum which is a combination of background and real source continuum.  Because of the short IRS high resolution slit, simultaneous background measurements cannot be subtracted from a single on-source observation so we do not determine the real source continuum before fitting the line.  Therefore, equivalent widths are not measured.  High resolution spectra are especially important for the [NeII] 12.81 \um line fluxes because an adjacent PAH feature makes accurate [NeII] measures impossible on low resolution spectra.  We treat the wings of the PAH as an underlying continuum when measuring the [NeII] flux.

Even after removal of known bad pixels, the IRS high resolution spectra are contaminated by noise spikes from individual ``rogue pixels".  These rarely interfere with individual emission lines at known wavelengths, although they are a serious contaminant when searching for weak, unknown emission features.   In those few cases when a single pixel noise spike interferes with an otherwise smooth emission line profile, such spikes are removed by visual inspection. This is done independently for each spectrum from the two nods of an observation, and line profiles are fit independently for each nod.  To estimate the uncertainty of line fluxes arising from the noise removal and from the line fitting process, line fluxes from the two spectra arising from the two nods were compared. As expected, line uncertainty is greater for fainter lines.  The dispersion of the ratio of measured fluxes between the two nods is $\pm$ 9\% for all lines, and $\pm$ 6\% for bright lines with log flux $>$ -20.2 (units of W cm$^{-2}$).  For quadratically combining IRS line flux uncertainty with the [CII] uncertainty of $\pm$ 20\%, we adopt the larger 9\% dispersion for all lines, giving a final uncertainty for flux ratios [CII]/IRS lines = $\pm$ 22\%.

\clearpage

\begin{deluxetable}{llcccccccccc} % 12 columns
\tablecolumns{12}
\rotate
\tabletypesize{\scriptsize}

\tablewidth{0pc}
\tablecaption{[CII] Line Fluxes and Luminosities for New Sources}
\tablehead{
%first line of header  
 \colhead{No.} &\colhead{Name}& \colhead{coordinates} &\colhead{z\tablenotemark{a}} & \colhead{EW\tablenotemark{b}} & \colhead{f[CII]\tablenotemark{c}}& \colhead{S/N\tablenotemark{d}} & \colhead{[CII]/11.3\tablenotemark{e}}  & \colhead{[CII]/7.8\tablenotemark{f}} & \colhead {$L([CII])$\tablenotemark{g}} & \colhead {$L([CII])$/$L_{ir}$\tablenotemark{h}} & \colhead {$Herschel$ id} \\
%second line of header
\colhead{} & \colhead{} & \colhead{} & \colhead{} & \colhead{6.2 \um} & \colhead{} & \colhead{}   & \colhead{}  & \colhead{} & \colhead{solar} &\colhead{} & \colhead {}\\
%third line of header
\colhead{} & \colhead{} & \colhead{J2000} & \colhead{} & \colhead{\um} & \colhead{W m$^{-2}$} & \colhead{}  & \colhead{log} & \colhead{log} & \colhead{log} & \colhead{log} & \colhead {} 
}
\startdata

1 & NGC1808 & 050742.32-373045.7 & 0.00337 & 0.57 & 1.21e-14 & 256.1 & -0.28 & -2.14 & 7.87 & -2.88 & 1342270686 \\
2 & IRASF07260+3955 & 072930.29+394941.0 & 0.07888 & 0.03 & 4.70e-17 & 22.5 & 0.14 & -2.42 & 8.25 & -3.34 & 1342251036 \\
3 & IRASF08076+3658 & 081056.13+364945.0 & 0.02156 & 0.51 & 3.16e-16 & 101.1 & 0.00 & -1.83 & 7.92 & -2.57 & 1342251347 \\
4 & IRASF08273+5543 & 083109.29+553316.0 & 0.04475 & 0.30 & 2.13e-17 & 6.0 & -0.56 & -2.47 & 7.39 & -3.80 & 1342270666 \\
5 & IRASF08344+5105 & 083803.63+505509.0 & 0.09688 & 0.36 & 8.87e-17 & 29.7 & -0.26 & -2.36 & 8.71 & -3.30 & 1342270665 \\
6 & IRASF09168+3308 & 091954.54+325559.0 & 0.10376\tablenotemark{i} & 0.48 & 4.37e-16\tablenotemark{i}  & 83.7 & -0.07 & -2.00 & 8.81 & -2.76 & 1342270660\\
7 & IRASF09414+4843 & 094442.24+482916.0 & 0.05500 & 0.06 & 7.03e-17 & 21.5 & -0.19 & -2.41 & 8.10 & -3.11 & 1342270663 \\
8 & IRASF09471+3158 & 095004.02+314442.0 & 0.01640 & 0.55 & 1.70e-16 & 69.5 & -0.49 & -2.33 & 7.41 & -2.86 & 1342271043 \\
9 & Mrk25 & 100351.90+592610.0 & 0.01007 & 0.45 & 2.47e-16 & 88.7 & -0.23 & -2.05 & 7.14 & -2.73 & 1342270667 \\
10 & IRASF10332+6338 & 103636.12+632222.0 & 0.03812 & 0.51 & 9.89e-17 & 34.2 & -0.38 & -2.29 & 7.92 & -2.96 & 1342270668 \\
11 & IRASF10590+6515 & 110213.02+645924.0 & 0.07777 & 0.02 & 4.69e-17 & 16.0 & -0.46 & -2.58 & 8.23 & -3.23 & 1342270669 \\
12 & Mrk206 & 122417.00+672624.0 & 0.00442 & 0.50 & 2.63e-16 & 87.5 & -0.09 & -1.93 & 6.46 & -2.68 & 1342270670 \\
13 & IRASF12538+6352 & 125554.00+633644.0 & 0.00943 & 0.58 & 5.54e-16 & 119.6 & -0.08 & -2.22 & 7.42 & -2.73 & 1342270671 \\
14 & IRASF13007+6405 & 130239.21+634929.0 & 0.04165 & 0.45 & 3.24e-17 & 11.0 & -0.50 & -2.36 & 7.52 & -3.29 & 1342270672 \\
15 & IRAS16487+5447 & 164946.88+544235.4 & 0.10376\tablenotemark{i} & 0.29 & 1.18e-16\tablenotemark{i} & 34.2 & -0.07 & -2.14 & 8.90 & -3.28 & 1342270675 \\
16 & IRAS16569+8105 & 165236.82+810016.6 & 0.04927 & 0.59 & 4.58e-16 & 98.9 & -0.15 & -2.00 & 8.81 & -2.59 & 1342270674 \\
17 & IRAS17028+5817 & 170341.91+581344.4 & 0.10608 & 0.38 & 1.52e-16 & 26.3 & -0.09 & -2.14 & 9.03 & -3.13 & 1342270676 \\
18 & IRAS18580+6527 & 185813.90+653124.0 & 0.17630 & 0.35 & 7.54e-17 & 11.2 & -0.36 & -2.20 & 9.19 & -3.02 & 1342270679 \\

%Sept1: flux changed for number 18.  see earlier versions for old numbers.  new my sleeping astronomer

%also here are the new numbers for the object 1342270679 EW = 0.67 flux of CII is 7.54*10^(-17) the redshift is 0.17630 the L CII continuum is 45.52 the L CII is 42.77 or 9.19 in solar the log L(CII)/L(IR) = -3.02 the s/n is 11.17 the log f(CII)/f(11.3) is = -0.36 the log f(CII)/f(7.8) is = -2.20

\enddata
\tablenotetext{a}{Redshift from [CII] emission line adopting rest wavelength of 157.741 \um using central velocity of Gaussian profile in brightest spaxel, corrected to the Local Standard of Rest (LSR) as in the HIPE pipeline, except for sources noted by footnote i with asymmetric or component structure for which central velocity at 50\% half maximum measured without applying Gaussian fit.}
\tablenotetext{b}{Rest frame equivalent width of 6.2 \um PAH emission feature from \citet{sar11} used to classify source as AGN, composite, or starburst as shown in Figures.}
\tablenotetext{c}{Total flux of [CII] 158 \um emission line in units of W m$^{-2}$ using Gaussian fit to line for simple profiles and integrated flux of line for complex profiles (noted by footnote i).  Line flux listed is the total flux observed within the 9 spaxels centered on the brightest spaxel, increased by a correction factor of 1.16 to 1.19 (depending on redshift) to include flux that would fall outside these spaxels for an unresolved source. The correction factor adopted for the range of observed [CII] wavelengths from 160 \um to 185 \um is 1.16($\lambda$/158 \ums)$^{0.17}$ using calibration communicated to us by the PACS calibration team.  Uncertainties of individual fits given by S/N in next column; systematic uncertainty for all fluxes depends on PACS flux calibration, estimated as $\pm$ 12\% in the PACS Spectroscopy Performance and Calibration document PICC-KL-TN-041.  All fluxes given in this table were extracted using HIPE v8 to be consistent with fluxes in Paper 1, but fluxes were also determined with HIPE v10 for which we note a systematic difference by a factor of 1.07 for flux ratios v8/v10.}
\tablenotetext{d}{Signal to noise ratio of total line flux in brightest spaxel, using one sigma uncertainty of profile fit.}
\tablenotetext{e}{Ratio of flux in [CII] 158 \um to flux of PAH 11.3 \um feature from \citet{sar11}.}
\tablenotetext{f}{Ratio f([CII])/$\nu$f$_{\nu}$(7.8 \ums) using f$_{\nu}$(7.8 \ums) from \citet{sar11}.}
\tablenotetext{g}{ [CII] 158 \um emission line luminosity $L([CII])$ in \ldot~using luminosity distances determined for H$_0$ = 71 \kmsMpc, $\Omega_{M}$=0.27 and $\Omega_{\Lambda}$=0.73, from \citet{wri06}:  http://www.astro.ucla.edu/~wright/CosmoCalc.html; [log $L([CII])$ (solar) = log $L([CII])$ (W) - 26.59]. }
\tablenotetext{h}{Ratio of [CII] luminosity to $L_{ir}$ using $L_{ir}$ given in \citet{sar11} from f$_{ir}$ determined as in \citet{san96}, f$_{ir}$ =  1.8 x 10$^{-11}$[13.48f$_{\nu}$(12) + 5.16f$_{\nu}$(25) + 2.58f$_{\nu}$(60) + f$_{\nu}$(100)] in erg cm$^{-2}$ s$^{-1}$ using IRAS flux densities at 12 \ums, 25 \ums, 60 \um and 100 \ums.}
\tablenotetext{i}{ [CII] line profile is asymmetric or has component structure so total line flux is the integrated flux including all components rather than flux within a single Gaussian fit.}

\end{deluxetable}

\clearpage

\begin{deluxetable}{llcccccccc} % 10 columns
\tablecolumns{10}
\rotate
\tabletypesize{\scriptsize}

\tablewidth{0pc}
\tablecaption{[CII] and Mid-Infrared Line Fluxes}
\tablehead{
%first line of header  
 \colhead{No.} &\colhead{Name\tablenotemark{a}} &\colhead{redshift\tablenotemark{b}} & \colhead{f[CII]\tablenotemark{c}}&\colhead{CII/NeII\tablenotemark{d}}&\colhead{CII/NeIII\tablenotemark{e}} &\colhead{CII/SIII\tablenotemark{f}}&\colhead{CII/SIV\tablenotemark{g}}&\colhead{CII/OIV\tablenotemark{h}}& \colhead{EW([CII])\tablenotemark{i}}\\ 

%second line of header
\colhead{} & \colhead{} & \colhead{} & \colhead{W m$^{-2}$} & \colhead{log} & \colhead{log} & \colhead{log}   & \colhead{log}  & \colhead{log} & \colhead{\um}
}

\startdata

%June 19; redshifts of asymmetric sources corrected to center of profile

1 & Mrk0334 & 0.02214 & 6.70e-16 & 0.22 & 0.60 & 0.53 & 1.36 & 0.85 & 1.84 $\pm$ 0.30 \\
2 & MCG-02-01-051/2 & 0.02724 & 1.18e-15 & 0.11 & 0.85 & 0.55 & 1.64 & 1.73 & 1.71 $\pm$ 0.16 \\
3 & IRAS00199-7426 & 0.09609 & 1.98e-16 & 0.00 & 1.03 & 0.5 & $>$1.76 & $>$1.18 & 0.47 $\pm$ 0.09 \\
4 & E12-G21 & 0.03295 & 5.76e-16 & 0.68 & 1.08 & 1.13 & 1.48 & 0.58 & 1.93 $\pm$ 0.41 \\
5 & IRASF00456-2904SW & 0.10994 & 1.26e-16 & 0.14 & 0.89 & 0.18 & $>$1.93 & $>$1.14 & 0.53 $\pm$ 0.15 \\
6 & MCG-03-04-014 & 0.03519 & 1.38e-15 & 0.31 & 1.18 & 0.73 & $>$2.04 & 1.30 & 1.52 $\pm$ 0.15 \\
7 & NGC0454 & 0.01214 & 1.55e-16 & 0.54 & 0.39 & $>$1.26 & 0.99 & 0.08 & 1.21 $\pm$ 1.46 \\
8 & ESO244-G012 & 0.02253 & 1.14e-15 & 0.01 & 0.68 & 0.39 & 1.72 & 1.43 & 1.26 $\pm$ 0.15 \\
9 & ESO353-G020 & 0.01607 & 1.56e-15 & 0.31 & 1.37 & 1.11 & 2.35 & $>$1.91 & 0.96 $\pm$ 0.04 \\
10 & IRASF01364-1042 & 0.04840 & 1.32e-16 & 0.22 & $>$1.56 & 1.05 & $>$1.74 & $>$1.29 & 0.32 $\pm$ 0.06 \\
11 & UGC01385 & 0.01848 & 5.70e-16 & 0.10 & 1.03 & 0.28 & $>$2.09 & $>$1.29 & 0.86 $\pm$ 0.07 \\
12 & NGC0788 & 0.01353 & 8.43e-17 & 0.14 & -0.19 & 0.20 & 0.09 & -0.42 & 1.01 $\pm$ 0.19 \\
13 & IRAS02054+0835 & 0.34499 & $<$2.00e-17 & $<$0.25 & \nodata & \nodata & $<$0.52 & \nodata & -0.03 $\pm$ 0.29 \\
14 & Mrk0590 & 0.02659 & 2.06e-16 & 0.42 & 0.94 & $>$1.31 & 1.29 & 0.81 & 1.03 $\pm$ 0.86 \\
15 & UGC01845 & 0.01570\tablenotemark{k} & 1.42e-15 & 0.20 & 1.20 & 0.87 & $>$2.26 & $>$2.08 & 0.86 $\pm$ 0.04 \\
16 & IC1816 & 0.01693 & 3.42e-16 & 0.37 & 0.26 & 0.55 & 0.76 & 0.39 & 1.07 $\pm$ 0.39 \\
17 & NGC0973 & 0.01619\tablenotemark{k} & 2.46e-16 & 0.53 & 0.46 & 0.67 & 1.04 & 0.30 & 0.67 $\pm$ 0.23 \\
18 & IRASF02437+2122 & 0.02331 & 1.50e-16 & -0.09 & 0.98 & 0.76 & $>$1.79 & $>$1.18 & 0.24 $\pm$ 0.03 \\
19 & UGC02369 & 0.03161\tablenotemark{k} & 8.21e-16 & 0.23 & 1.15 & 0.69 & 1.67 & $>$2.21 & 0.97 $\pm$ 0.12\\
20 & Mrk1066 & 0.01216 & 9.39e-16 & -0.03 & 0.31 & 0.32 & 1.00 & 0.45 & 0.90 $\pm$ 0.06 \\
21 & IRASF03217+4022 & 0.02350\tablenotemark{k} & 5.96e-16 & 0.27 & 1.25 & 0.95 & $>$2.03 & $>$1.66 & 0.61 $\pm$ 0.04 \\
22 & Mrk0609 & 0.03460 & 5.44e-16 & 0.47 & 0.90 & 0.85 & $>$1.72 & $>$1.66 & 1.46 $\pm$ 0.20 \\
23 & IRASF03359+1523 & 0.03566 & 4.60e-16 & 0.17 & 0.68 & 0.54 & 1.45 & $>$1.29 & 0.87 $\pm$ 0.18 \\
24 & IRASF03450+0055 & 0.03079 & 2.69e-17 & 0.42 & 0.03 & $>$0.33 & 0.13 & -0.04 & 0.78 $\pm$ 1.57 \\
25 & IRAS03538-6432 & 0.30069 & $<$2.00e-17 & $<$-0.28 & \nodata & \nodata & \nodata & \nodata & 0.10 $\pm$ 0.14 \\
26 & IRAS04103-2838 & 0.11773 & 8.53e-17 & -0.07 & 0.05 & 0.41 & 0.78 & 0.25 & 0.77 $\pm$ 0.51 \\
27 & IRAS04114-5117 & 0.12512 & 6.32e-17 & \nodata & \nodata & $>$0.66 & \nodata & $>$0.99 & 0.46 $\pm$ 0.58 \\
28 & ESO420-G013 & 0.01207 & 1.38e-15 & 0.07 & 0.73 & 0.50 & 1.27 & 0.43 & 0.63 $\pm$ 0.02 \\
29 & 3C120 & 0.03315 & 3.62e-16 & 0.68 & 0.12 & 0.57 & 0.19 & -0.48 & 1.78 $\pm$ 0.58 \\
30 & ESO203-IG001 & 0.05293 & 5.14e-17 & 0.50 & \nodata & $>$0.78 & $>$0.98 & $>$0.41 & 0.19 $\pm$ 0.04 \\
31 & MCG-05-12-006 & 0.01852 & 3.74e-16 & -0.11 & 0.96 & 0.29 & $>$1.90 & $>$1.24 & 0.43 $\pm$ 0.03 \\
32 & NGC1808\tablenotemark{j} & 0.00337 & 1.21e-14 & \nodata & \nodata & \nodata & \nodata & \nodata & 0.96 $\pm$ 0.02 \\
33 & Ark120 & 0.03285 & 1.34e-16 & 0.77 & 0.76 & $>$0.89 & 0.93 & 0.58 & 0.78 $\pm$ 0.48 \\
34 & VIIZw31 & 0.05430 & 8.55e-16 & 0.22 & 1.06 & 0.67 & $>$2.21 & \nodata & 1.02 $\pm$ 0.07 \\
35 & IRASF05187-1017 & 0.02861 & 2.20e-16 & 0.30 & 1.06 & 1.02 & $>$1.91 & $>$1.44 & 0.29 $\pm$ 0.03 \\
36 & 2MASXJ05580206-3820043 & 0.03421 & 1.29e-17 & -0.27 & -0.61 & -0.60 & $>$-0.15 & -0.59 & 0.58 $\pm$ 0.98 \\
37 & IRASF06076-2139 & 0.03763 & 1.11e-16 & -0.14 & $>$1.28 & 0.66 & $>$1.22 & $>$0.86 & 0.19 $\pm$ 0.02 \\
38 & IRAS06301-7934 & 0.15632 & 3.10e-17 & 0.34 & \nodata & $>$0.78 & $>$0.88 & $>$0.49 & 0.16 $\pm$ 0.10 \\
39 & IRAS06361-6217 & 0.15997 & 3.38e-17 & -0.03 & 0.55 & $>$0.88 & 0.59 & 0.18 & 0.41 $\pm$ 0.78 \\
40 & NGC2273 & 0.00618 & 7.18e-16 & 0.25 & 0.58 & 0.70 & 1.20 & 0.71 & 1.05 $\pm$ 0.10 \\
41 & UGC03608 & 0.02163 & 1.46e-15 & 0.50 & 1.30 & 0.75 & 2.26 & $>$2.22 & 1.15 $\pm$ 0.11 \\
42 & IRASF06592-6313 & 0.02304 & 2.51e-16 & -0.12 & 0.83 & 0.47 & $>$1.92 & $>$1.60 & 0.41 $\pm$ 0.04 \\
43 & IRASF07260+3955\tablenotemark{j} & 0.07887 & 4.70e-17 & \nodata & \nodata & \nodata & \nodata & \nodata & 1.63 $\pm$ 3.55 \\
44 & Mrk0009 & 0.04009 & 1.12e-16 & 0.59 & 0.65 & $>$1.08 & 0.55 & 0.36 & 0.76 $\pm$ 0.32 \\
45 & IRAS07598+6508 & 0.14866 & 2.87e-17 & -0.06 & 0.12 & $>$0.95 & $>$1.34 & $>$0.52 & 0.21 $\pm$ 0.17 \\
46 & Mrk0622 & 0.02351 & 8.36e-17 & -0.06 & 0.22 & 0.33 & 0.66 & $>$0.68 & 0.48 $\pm$ 0.17 \\
47 & IRASF08076+3658\tablenotemark{j} & 0.02155 & 3.16e-16 & \nodata & \nodata & \nodata & \nodata & \nodata & 2.26 $\pm$ 1.10 \\
48 & IRASF08273+5543\tablenotemark{j} & 0.04474 & 2.13e-17 & \nodata & \nodata & \nodata & \nodata & \nodata & 0.09 $\pm$ 0.09 \\
49 & IRASF08344+5105\tablenotemark{j} & 0.09688 & 8.87e-17 & \nodata & \nodata & \nodata & \nodata & \nodata & 0.59 $\pm$ 0.25 \\
50 & ESO60-IG016 & 0.04488 & 4.99e-16 & 0.36 & 0.44 & 0.60 & \nodata & $>$1.32 & 1.26 $\pm$ 0.24 \\
51 & Mrk0018 & 0.01133 & 6.47e-16 & 0.43 & 0.88 & 0.86 & $>$1.95 & $>$1.63 & 2.64 $\pm$ 0.80 \\
52 & IRASF09168+3308\tablenotemark{j} & 0.10346\tablenotemark{k} & 4.37e-16 & \nodata & \nodata & \nodata & \nodata & \nodata & 1.04 $\pm$ 0.36 \\
53 & MCG-01-24-012 & 0.01981 & 1.39e-16 & 0.36 & 0.38 & 0.36 & 0.81 & 0.16 & 0.80 $\pm$ 0.43 \\
54 & Mrk0705 & 0.02879 & 1.34e-16 & 0.41 & 0.41 & 0.69 & 0.86 & 0.28 & 1.15 $\pm$ 0.83 \\
55 & IRASF09414+4843\tablenotemark{j} & 0.05500 & 7.03e-17 & \nodata & \nodata & \nodata & \nodata & \nodata & 1.62 $\pm$ 4.43 \\
56 & IRASF09471+3158\tablenotemark{j} & 0.01639 & 1.70e-16 & \nodata & \nodata & \nodata & \nodata & \nodata & 1.02 $\pm$ 0.26 \\
57 & Mrk25\tablenotemark{j} & 0.01007 & 2.47e-16 & \nodata & \nodata & \nodata & \nodata & \nodata & 1.88 $\pm$ 0.65 \\
58 & IRASF10038-3338 & 0.03421 & 4.03e-16 & 0.36 & 0.89 & 0.62 & $>$1.86 & $>$1.48 & 0.74 $\pm$ 0.15 \\
59 & IRASF10332+6338\tablenotemark{j} & 0.03811 & 9.89e-17 & \nodata & \nodata & \nodata & \nodata & \nodata & 2.13 $\pm$ 1.80 \\
60 & NGC3393 & 0.01258 & 3.44e-16 & 0.30 & -0.25 & 0.01 & -0.19 & -0.77 & 1.88 $\pm$ 0.85 \\
61 & IRASF10590+6515\tablenotemark{j} & 0.07777 & 4.69e-17 & \nodata & \nodata & \nodata & \nodata & \nodata & 0.64 $\pm$ 0.96 \\
62 & IRAS11119+3257 & 0.18900 & $<$2.00e-17 & $<$-0.10 & $<$0.07 & $<$0.40 & \nodata & \nodata & 0.23 $\pm$ 0.44 \\
63 & ESO319-G022 & 0.01634 & 4.00e-16 & 0.22 & 1.35 & 0.85 & $>$1.61 & $>$1.26 & 0.29 $\pm$ 0.02 \\
64 & IRAS12018+1941 & 0.16809 & 2.41e-17 & -0.07 & 0.84 & $>$0.46 & $>$1.36 & $>$0.19 & 0.41 $\pm$ 0.44 \\
65 & UGC07064 & 0.02507 & 3.94e-16 & 0.40 & 0.78 & 0.99 & 1.03 & 0.54 & 0.88 $\pm$ 0.22 \\
66 & Mrk206\tablenotemark{j} & 0.00441 & 2.63e-16 & \nodata & \nodata & \nodata & \nodata & \nodata & 2.36 $\pm$ 0.93 \\
67 & NGC4507 & 0.01187 & 5.62e-16 & 0.25 & 0.28 & 0.55 & 0.78 & 0.20 & 1.41 $\pm$ 0.37 \\
68 & PG1244+026 & 0.04862 & 1.65e-17 & 0.02 & 0.11 & $>$0.25 & $>$0.49 & $>$0.08 & 1.10 $\pm$ 1.59 \\
69 & IRAS12514+1027 & 0.31820 & $<$2.00e-17 & $<$-0.04 & $<$-0.18 & \nodata & $<$0.33 & \nodata & 0.02 $\pm$ 0.11 \\
70 & IRASF12538+6352\tablenotemark{j} & 0.00942 & 5.54e-16 & \nodata & \nodata & \nodata & \nodata & \nodata & 1.13 $\pm$ 0.17 \\
71 & IRASF13007+6405\tablenotemark{j} & 0.04165 & 3.24e-17 & \nodata & \nodata & \nodata & \nodata & \nodata & 0.96 $\pm$ 1.49 \\
72 & ESO507-G070 & 0.02154 & 9.51e-16 & 0.33 & 1.11 & 0.99 & $>$2.34 & $>$1.74 & 0.81 $\pm$ 0.04 \\
73 & NGC4941 & 0.00372 & 1.71e-16 & 0.11 & -0.13 & 0.35 & 0.23 & -0.2 & 0.85 $\pm$ 0.25 \\
74 & ESO323-G077 & 0.01532 & 8.27e-16 & 0.33 & 0.69 & 0.71 & 1.02 & 0.52 & 1.02 $\pm$ 0.08 \\
75 & MCG-03-34-064 & 0.01682 & 1.90e-16 & -0.42 & -0.76 & -0.19 & -0.40 & -0.73 & 0.47 $\pm$ 0.09 \\
76 & IRASF13279+3401 & 0.02364 & 2.58e-17 & -0.34 & 1.09 & 0.80 & $>$1.28 & $>$1.26 & 0.22 $\pm$ 0.15 \\
77 & M-6-30-15 & 0.00800 & 7.04e-17 & 0.31 & 0.06 & -0.03 & 0.00 & \nodata & 0.75 $\pm$ 0.37 \\
78 & IRAS13342+3932 & 0.17962 & 6.54e-17 & 0.10 & 0.15 & 0.39 & 0.52 & -0.17 & 1.08 $\pm$ 3.53 \\
79 & IRAS13352+6402 & 0.23660 & $<$2.00e-17 & $<$-0.15 & \nodata & \nodata & \nodata & \nodata & 0.21 $\pm$ 0.30 \\
80 & IRASF13349+2438 & 0.10840 & 2.54e-17 & 0.06 & -0.01 & $>$0.94 & -0.26 & -0.41 & 0.41 $\pm$ 0.44 \\
81 & NGC5347 & 0.00813 & 1.15e-16 & 0.38 & 0.43 & 0.93 & 0.85 & 0.24 & 0.62 $\pm$ 0.24 \\
82 & IRAS14026+4341 & 0.32330 & $<$2.00e-17 & \nodata & \nodata & \nodata & \nodata & \nodata & -0.01 $\pm$ 0.07 \\
83 & OQ+208 & 0.07673 & 6.28e-17 & 0.20 & 0.21 & $>$0.77 & $>$1.15 & $>$0.55 & 0.99 $\pm$ 0.97 \\
84 & NGC5548 & 0.01726 & 1.70e-16 & 0.30 & 0.26 & 0.53 & 0.62 & 0.14 & 0.89 $\pm$ 0.31 \\
85 & Mrk1490 & 0.02595 & 3.18e-16 & -0.16 & 0.95 & 0.51 & $>$1.75 & $>$1.76 & 0.45 $\pm$ 0.03 \\
86 & PG1426+015 & 0.08633 & 2.30e-17 & 0.28 & 0.00 & 0.00 & 0.22 & -0.31 & 0.17 $\pm$ 0.19 \\
87 & PG1440+356 & 0.07769 & 5.77e-17 & 0.15 & 0.17 & 0.23 & 0.57 & 0.18 & 0.55 $\pm$ 0.40 \\
88 & NGC5728 & 0.00949 & 1.25e-15 & 0.57 & 0.36 & 0.72 & 0.59 & 0.04 & 1.10 $\pm$ 0.06 \\
89 & NGC5793 & 0.01164\tablenotemark{k} & 8.98e-16 & 0.23 & 0.99 & 0.91 & $>$2.14 & 1.07 & 0.89 $\pm$ 0.05 \\
90 & IRAS15001+1433 & 0.16236 & 6.40e-17 & 0.01 & 0.38 & 0.41 & 1.12 & $>$0.52 & 0.43 $\pm$ 0.38 \\
91 & IRAS15225+2350 & 0.13875 & 3.23e-17 & -0.04 & 0.52 & \nodata & 0.14 & \nodata & 0.42 $\pm$ 0.31 \\
92 & Mrk0876 & 0.12941 & 2.99e-17 & -0.09 & -0.00 & -0.15 & 0.35 & -0.28 & 0.28 $\pm$ 0.43 \\
93 & IRASF16164-0746 & 0.02350\tablenotemark{k} & 7.51e-16 & 0.19 & 0.74 & 1.07 & 1.84 & 1.02 & 0.65 $\pm$ 0.03 \\
94 & Mrk0883 & 0.03827 & 1.84e-16 & 0.19 & 0.37 & 0.45 & 1.00 & 0.28 & 1.46 $\pm$ 1.06 \\
95 & CGCG052-037 & 0.02478 & 1.13e-15 & 0.19 & 1.32 & 0.75 & $>$2.49 & $>$2.28 & 0.98 $\pm$ 0.06 \\
96 & IRAS16334+4630 & 0.19085 & 3.62e-17 & -0.16 & 0.51 & 0.32 & 0.55 & $>$0.81 & 0.49 $\pm$ 0.73 \\
97 & ESO069-IG006 & 0.04644\tablenotemark{k} & 1.08e-15 & 0.16 & 0.93 & 0.67 & $>$1.97 & $>$1.39 & 0.81 $\pm$ 0.04 \\
98 & IRASF16399-0937N & 0.02728 & 1.49e-15 & 0.86 & 1.38 & 1.53 & $>$2.49 & 1.47 & 1.27 $\pm$ 0.65 \\
99 & IRAS16487+5447\tablenotemark{j} & 0.10361\tablenotemark{k} & 11.7e-17 & \nodata & \nodata & \nodata & \nodata & \nodata & 0.67 $\pm$ 0.46 \\
100 & IRAS16569+8105\tablenotemark{j} & 0.04926 & 4.58e-16 & \nodata & \nodata & \nodata & \nodata & \nodata & 2.06 $\pm$ 0.68 \\
101 & 2MASSJ165939.77+183436.9 & 0.17069 & $<$2.00e-17 & $<$0.18 & $<$-0.17 & \nodata & $<$-0.04 & \nodata & 0.19 $\pm$ 0.39 \\
102 & PG1700+518 & 0.29200 & $<$2.00e-17 & $<$0.09 & \nodata & \nodata & \nodata & \nodata & -0.07 $\pm$ 0.27 \\
103 & IRAS17028+5817\tablenotemark{j} & 0.10608 & 1.52e-16 & \nodata & \nodata & \nodata & \nodata & \nodata & 0.54 $\pm$ 0.18 \\
104 & IRAS17044+6720 & 0.13530 & 2.06e-17 & -0.43 & 0.16 & \nodata & 0.37 & \nodata & 0.39 $\pm$ 0.35 \\
105 & IRAS17068+4027 & 0.17933 & 3.37e-17 & -0.19 & -0.01 & 0.10 & $>$1.11 & $>$0.78 & 0.24 $\pm$ 0.48 \\
106 & IRASF17132+5313 & 0.05110\tablenotemark{k} & 6.09e-16 & 0.21 & 1.03 & 0.64 & $>$2.19 & $>$1.94 & 1.45 $\pm$ 0.32 \\
107 & ESO138-G027 & 0.02093 & 1.44e-15 & 0.52 & 1.43 & 0.88 & $>$2.59 & $>$1.89 & 1.10 $\pm$ 0.08 \\
108 & CGCG141-034 & 0.02015 & 6.65e-16 & 0.25 & 0.99 & 0.75 & 2.07 & 1.18 & 0.79 $\pm$ 0.07 \\
109 & H1821+643 & 0.29700 & $<$2.00e-17 & $<$-0.21 & $<$-0.7 & $<$-0.18 & $<$-0.46 & $<$-1.04 & 0.22 $\pm$ 0.37 \\
110 & IC4734 & 0.01548 & 1.36e-15 & 0.34 & 1.39 & 0.96 & 2.18 & $>$2.15 & 0.43 $\pm$ 0.02 \\
111 & IRAS18443+7433 & 0.13417 & 2.31e-17 & -0.05 & 0.06 & $>$0.83 & $>$0.76 & $>$0.91 & 0.32 $\pm$ 0.45 \\
112 & ESO140-G043 & 0.01419 & 1.76e-16 & 0.28 & 0.111 & 0.31 & 0.33 & -0.12 & 1.04 $\pm$ 0.36 \\
113 & 1H1836-786 & 0.07478 & 2.49e-17 & 0.21 & 0.34 & $>$0.211 & 0.41 & -0.32 & 0.27 $\pm$ 0.18 \\
114 & IRAS18580+6527\tablenotemark{j} & 0.17630 & 7.54e-17 & \nodata & \nodata & \nodata & \nodata & \nodata & 0.67 $\pm$ 0.31 \\
115 & ESO593-IG008 & 0.04923\tablenotemark{k} & 9.51e-16 & 0.44 & 1.07 & 0.66 & 2.09 & $>$1.71 & 0.77 $\pm$ 0.06 \\
116 & ESO-141-G055 & 0.03725 & 1.32e-16 & 0.69 & 0.41 & $>$0.99 & 0.35 & 0.09 & 0.87 $\pm$ 0.40 \\
117 & ESO339-G011 & 0.01918 & 1.54e-15 & 0.56 & 0.56 & 0.85 & 1.33 & 0.54 & 1.08 $\pm$ 0.09 \\
118 & IRAS20037-1547 & 0.19219 & 4.63e-17 & -0.04 & $>$0.62 & 0.38 & $>$1.16 & $>$0.96 & 0.30 $\pm$ 0.32 \\
119 & NGC6860 & 0.01479 & 3.69e-16 & 0.89 & 0.73 & 0.97 & 0.96 & 0.52 & 1.58 $\pm$ 0.93 \\
120 & ESO286-G035 & 0.01748 & 1.63e-15 & 0.53 & 1.45 & 0.91 & $>$2.65 & $>$2.28 & 1.28 $\pm$ 0.07 \\
121 & NGC7213 & 0.00598 & 2.74e-16 & 0.06 & 0.33 & 0.94 & $>$1.69 & $>$1 & 0.87 $\pm$ 0.26 \\
122 & ESO602-G025 & 0.02529 & 1.24e-15 & 0.41 & 1.22 & 1.04 & 2.15 & 1.37 & 1.39 $\pm$ 0.10 \\
123 & UGC12138 & 0.02516 & 1.43e-16 & 0.08 & 0.41 & $>$1.23 & 0.79 & 0.16 & 1.50 $\pm$ 1.67 \\
124 & UGC12150 & 0.02171 & 8.96e-16 & 0.24 & 1.37 & 0.77 & $>$2.31 & $>$1.86 & 0.56 $\pm$ 0.03 \\
125 & ESO239-IG002 & 0.04299 & 1.91e-16 & -0.14 & 0.76 & $>$1.48 & $>$1.71 & $>$1.34 & 0.31 $\pm$ 0.04 \\
126 & Zw453.062 & 0.02493 & 1.05e-15 & 0.61 & 1.20 & 1.24 & 2.27 & 1.27 & 0.90 $\pm$ 0.07 \\
127 & IRAS23060+0505 & 0.17300 & $<$2.00e-17 & $<$-0.17 & $<$-0.14 & $<$0.19 & $<$-0.02 & $<$-0.17 & 0.17 $\pm$ 0.24 \\
128 & NGC7603 & 0.02952\tablenotemark{k} & 2.63e-16 & 0.36 & 0.73 & 0.69 & $>$1.33 & 0.98 & 0.76 $\pm$ 0.12 \\
129 & MCG-83-1 & 0.02317 & 1.47e-15 & 0.44 & 1.17 & 0.87 & $>$2.15 & $>$2.33 & 1.71 $\pm$ 0.22 \\
130 & CGCG381-051 & 0.03091 & 2.34e-16 & 0.16 & 1.01 & 0.65 & $>$1.49 & $>$1.29 & 0.66 $\pm$ 0.21 \\
%sept1; number changed for 114, was 4.64 for flux, was 0.27 for EW, and was 0.17626 for z.  changed because needed to fit with broadened profile after Lusi checked all based on new fits including PSF
\enddata

\tablenotetext{a}{Source name as in \citet{sar12} or Table 1 (sources with note j) where coordinates, PAH 11.3 \um fluxes, and PAH 6.2 \um EW are listed. }
\tablenotetext{b}{Redshift from [CII] emission line adopting rest wavelength of 157.741 \um using central velocity of Gaussian profile fit in brightest spaxel, corrected to LSR in the HIPE pipeline. For objects noted by footnote k having asymmetric profiles or component structure, the [CII] redshift is the central velocity at half maximum of the profile in the brightest spaxel without applying Gaussian fit. }
\tablenotetext{c}{Flux of [CII] 158 \um emission line in units of W m$^{-2}$ from \citet{sar12} or from Table 1 which includes new sources from program lsargsyan-OT2.}
\tablenotetext{d}{Ratio of flux in [CII] 158 \um emission line to flux of [NeII] 12.81 \um emission line, fit as a Gaussian in IRS high resolution spectrum. } 
\tablenotetext{e}{Ratio of flux in [CII] 158 \um emission line to flux of [NeIII] 15.55 \um emission line, fit as a Gaussian in IRS high resolution spectrum.} 
\tablenotetext{f}{Ratio of flux in [CII] 158 \um emission line to flux of [SIII] 18.71 \um emission line, fit as a Gaussian in IRS high resolution spectrum.} 
\tablenotetext{g}{Ratio of flux in [CII] 158 \um emission line to flux of [SIV] 10.51 \um emission line, fit as a Gaussian in IRS high resolution spectrum.} 
\tablenotetext{h}{Ratio of flux in [CII] 158 \um emission line to flux of [OIV] 25.89 \um emission line, fit as a Gaussian in IRS high resolution spectrum. } 
\tablenotetext{i}{Rest frame equivalent width in \um of [CII] emission line.  Uncertainties are quadratically combined 1 $\sigma$ uncertainties from line flux measurement and from measurement of continuum level.}
\tablenotetext{j} {Source listed in Table 1.}
\tablenotetext{k} {[CII] profile is asymmetric or shows component structure so total line flux is integrated flux including all components rather than flux within a single Gaussian fit.}

\end{deluxetable}

\section{Discussion}

\subsection {Comparison of [CII] Fluxes and Mid-Infrared Emission Lines}

\begin{figure}
\figurenum{3}
\includegraphics[scale=0.9]{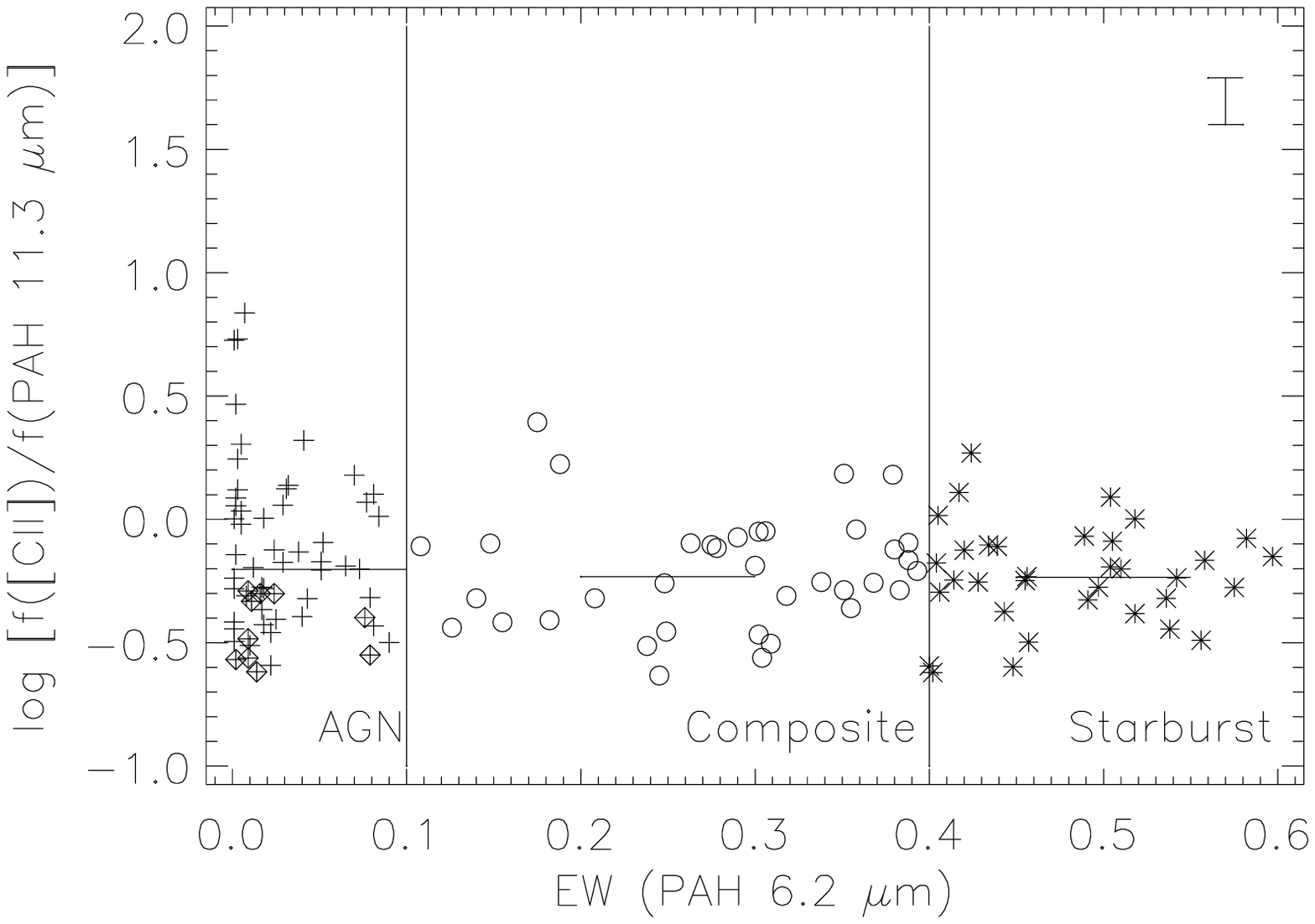}
\caption{Ratio of [CII] 158 \um to PAH 11.3 \um line fluxes, compared to source classification from EW(PAH 6.2 \ums) measured in \ums, for all sources.  Crosses are AGN from the EW(PAH 6.2 \ums) classification, open circles are composite AGN plus starburst, and asterisks are starbursts.  Sources with diamonds (all AGN) are upper limits to [CII] line fluxes.  Horizontal bars are medians within each category; medians include limits because all limits fall below the median.  Vertical error bar shows the observational line ratio uncertainty for individual points of $\pm$ 22\% derived in the text. }  
\end{figure}

\begin{figure}
\figurenum{4}
\includegraphics[scale=0.9]{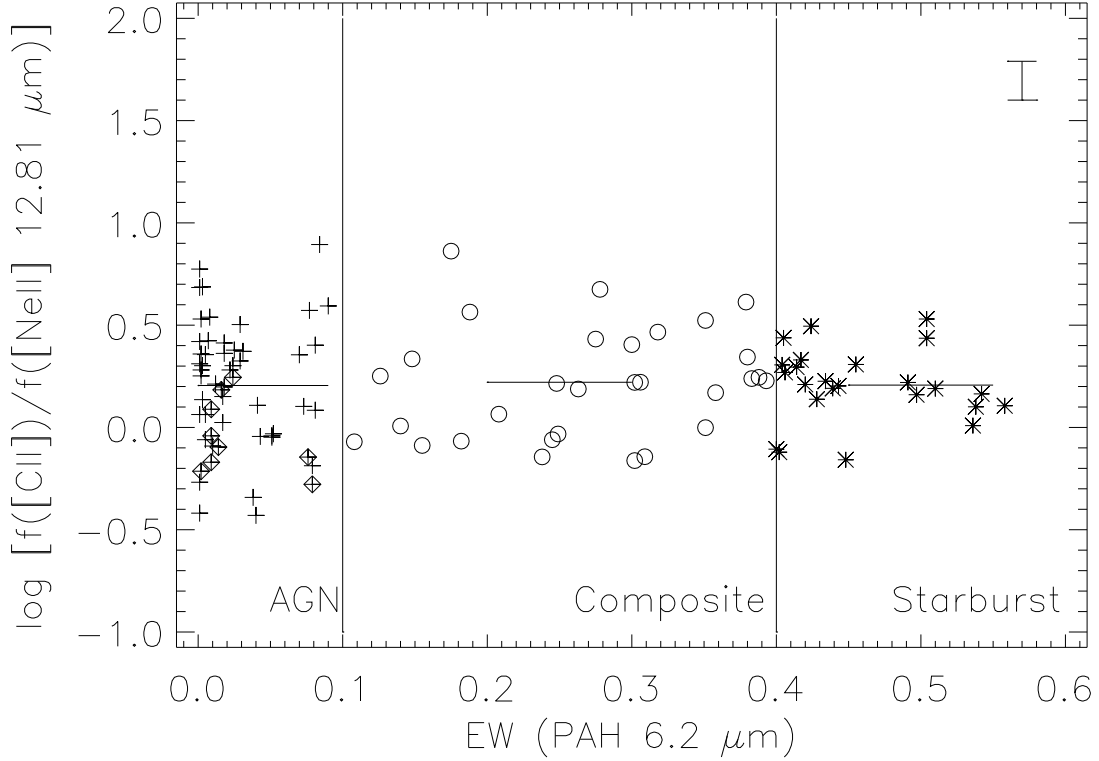}
\caption{Ratio of [CII] 158 \um to [NeII] 12.81 \um line fluxes, compared to source classification from EW(PAH 6.2 \ums) measured in \ums.  Crosses are AGN from the EW(PAH 6.2 \ums) classification, open circles are composite AGN plus starburst, and asterisks are starbursts.  Sources with diamonds (all AGN) are upper limits to ratio because are limits to [CII] line fluxes but detections in [NeII].  Horizontal bars are medians within each category, including limits.    Vertical error bar shows the observational line ratio uncertainty for individual points of $\pm$ 22\% derived in the text. 
 }    
\end{figure}

\begin{figure}
\figurenum{5}
\includegraphics[scale=0.9]{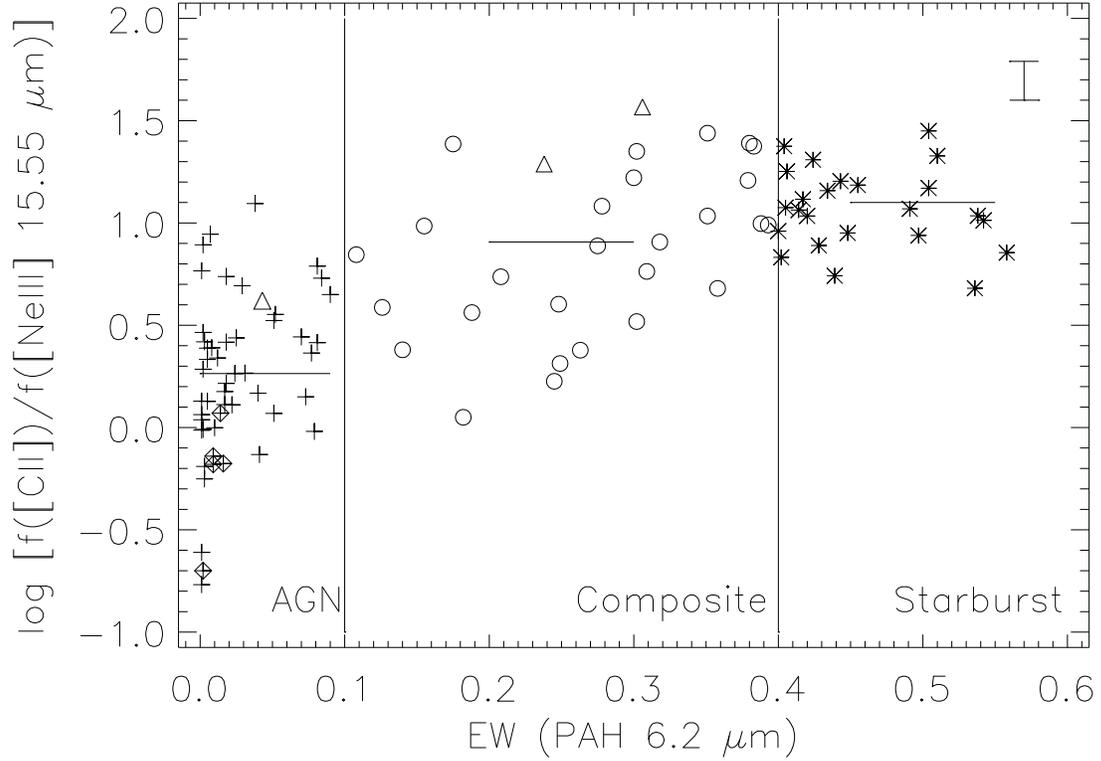}
\caption{Ratio of [CII] 158 \um to [NeIII] 15.55 \um line fluxes, compared to source classification from EW(PAH 6.2 \ums) measured in \ums.  Crosses are AGN from the EW(PAH 6.2 \ums) classification, open circles are composite AGN plus starburst, and asterisks are starbursts.  Sources with diamonds (all AGN) are upper limits to ratio because are limits to [CII] line fluxes but detections in [NeIII].  Triangles are lower limits to the ratio because values arise from limits for [NeIII] but detections in [CII].  Horizontal bars are medians within each category, including limits.  Vertical error bar shows the observational line ratio uncertainty for individual points of $\pm$ 22\% derived in the text. }  
\end{figure}

\begin{figure}
\figurenum{6}
\includegraphics[scale=0.9]{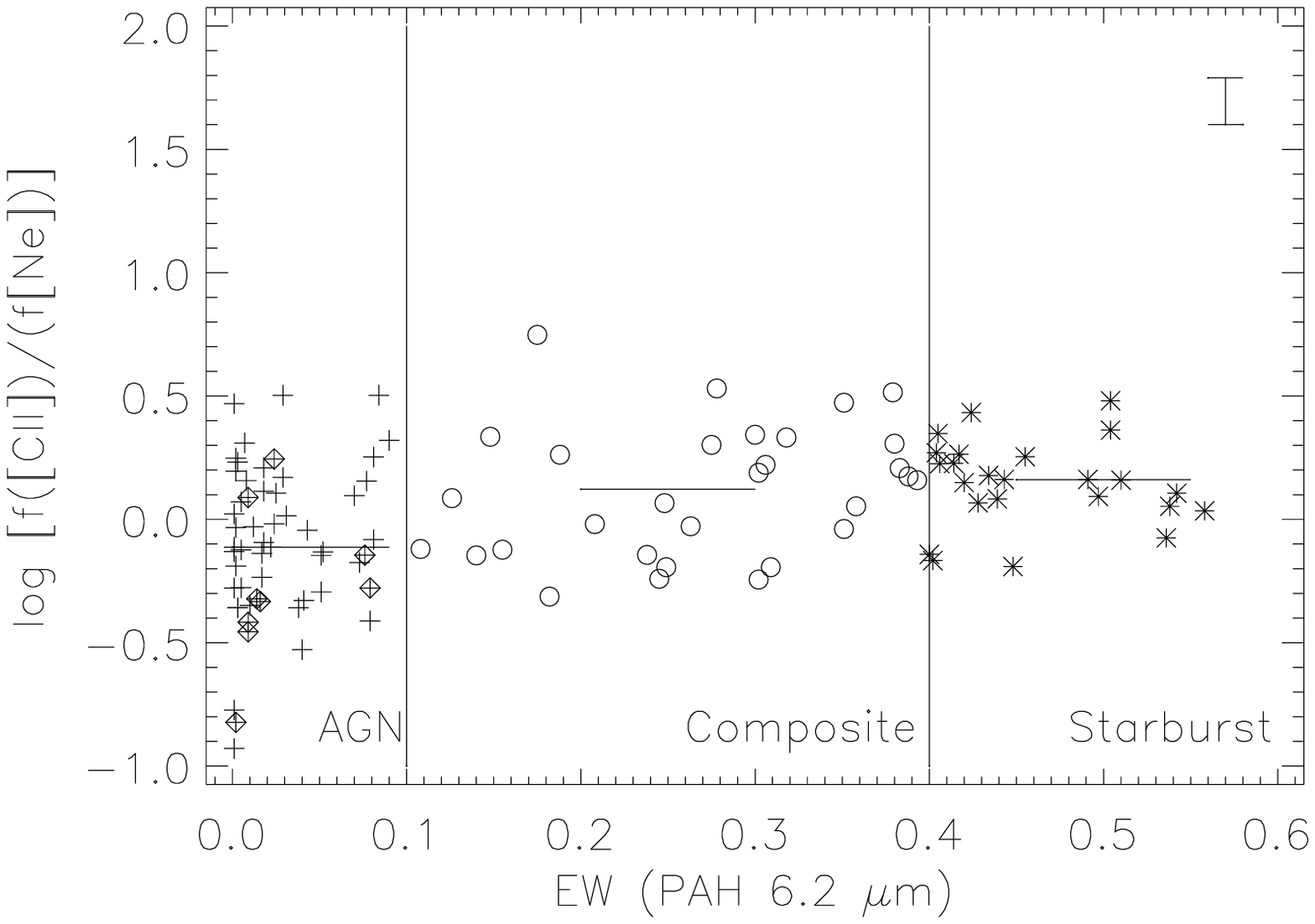}
\caption{ Ratio of [CII] 158 \um to [NeII] 12.81 \um + [NeIII] 15.55 \um line fluxes, compared to source classification from EW(PAH 6.2 \ums) measured in \ums.  Crosses are AGN from the EW(PAH 6.2 \ums) classification, open circles are composite AGN plus starburst, and asterisks are starbursts.  Sources with diamonds are upper limits to the ratio because values are limits to [CII] line fluxes but detections in [NeIII] and/or [NeII].  Horizontal bars are medians within each category, including limits.  Vertical error bar shows the observational line ratio uncertainty for individual points of $\pm$ 22\% derived in the text.  }  
\end{figure}

In Figures 3-6, the [CII] flux is compared with mid-infrared features from IRS spectra.  All scales are the same so ratios can be compared among the plots.  Figure 3 repeats the comparison with PAH 11.3 \um from Paper 1, adding the new sources in Table 1.  Figure 4 compares [CII] to [NeII], Figure 5 [CII] to [NeIII], and Figure 6 [CII] to [NeII]+[NeIII].

The comparison with PAH in Figure 3 reconfirms the conclusion in Paper 1, that the scaling between [CII] and PAH is independent of source classification, implying that the starburst component and SFR is measured equally well by PAH emission or by [CII] emission in sources of all classifications.  The correlation of [CII] with a SFR indicator can be checked in a completely independent way using Figure 4, which compares [CII] with [NeII], because [NeII] is a luminosity indicator of star formation determined by the HII region instead of the surrounding PDR. 

Figure 4 shows a result very similar to the [CII]/PAH comparison.  The median ratios f([CII])/f([NeII]) are independent of classification.  The dispersion about the median is also similar to the dispersion in [CII]/PAH.  Because [NeII] arises primarily from starbursts, this result is very important empirically because it confirms the previous conclusion from PAH that [CII] scales with the starburst component within sources of all classifications.    

%Another important use of this figure will be to consider the sources with largest ratios [CII]/[NeII] in context of looking for very obscured starbursts.  Are any of these same as sources with largest [CII]/PAH???   Note that, once again, most extreme are among the AGN - implying an additional [CII] component for some AGN.  But is also consistent with obscured starbursts having weak [NeII] and also having a weak PAH that gives the AGN classification.  In this case, also expect weak [OIV] if misclassified - can check that for individual sources.

Figure 5 compares [CII] with [NeIII].  There is a greater difference of ratio with classification than for [NeII].   The increasing ratio f([CII])/f([NeIII]) from AGN to starbursts can be explained if [NeIII] arises both from AGN ionization and starburst ionization, but [CII] arises only from starbursts, so that AGN contain an additional [NeIII] component compared to [CII].  The implications  of these results are discussed in the next section.

\subsection {Star Formation Rate from [CII] Calibrated with Neon Emission Lines}

The [CII] line is expected to arise primarily within the photodissociation region (PDR) surrounding star forming regions \citep{tie85,hel01,mal01,mei07}.  As the primary cooling line for low ionization regions, [CII] can also arise within more diffuse emission regions associated with the ``infrared cirrus", for which the ionization and heating of CII (and the accompanying far infrared dust continuum) is not necessarily a measure of the young, ongoing starbursts.  This diffuse [CII] is observed in nearby, resolved galaxies \citep{ken11}.  The scaling of [CII] with PAH found in Paper 1 for luminous, unresolved sources indicates, however, that [CII] primarily measures the PDRs in spatially integrated observations that include entire galaxies rather than individual regions within galaxies.  

In paper 1,  SFR was determined by comparison of [CII] to total infrared luminosity, using the assumption \citep{ken98} that all of the radiation from ongoing star formation is absorbed and reemitted in the infrared spectrum by the dust.  A completely independent estimate of SFR can be made by ionization models of the HII region, where the emission lines are related only to the ultraviolet ionizing continuum of the starburst.  A great advantage of using mid-infrared emission lines for comparison to the models is that they are much less subject to uncertainties regarding extinction corrections compared to optical lines. 

Starburst models have been scaled to the luminosity of the infrared [NeII] and [NeIII] lines by \citet{ho07} and discussed in comparison with $Spitzer$ IRS observations by \citet{far07}.  The Ho and Keto result is SFR (\mdot) = 4.34 x 10$^{-41}$ L([NeII]+[NeIII]) (erg s$^{-1}$ ), estimating about a factor of two uncertainty for individual sources depending on the fraction of ionizing photons which are absorbed by gas and the fractional ionizing state of Neon.  These differences arising from cosmic variance among sources can explain the dispersion in the [CII]/([NeII]+[NeIII]) ratio in Figure 6, for example.  In what follows, we use only the pure starbursts in Figure 6 for comparisons to avoid any uncertainty regarding Neon luminosity from an AGN component. 

Figure 6 shows the scaling with [CII].   For the pure starbursts in our sample, the result is log f[CII]/(f[NeII]+Ne[III]) = 0.16 $\pm$ 0.15.  Composites give a similar result, but, as expected from the individual plots for these lines in Figures 4 and 5, the value for AGN is lower.  This scaling for pure starbursts applied to the Ho and Keto relation gives the result that SFR (\mdot) = 3 x 10$^{-41}$ $L([CII])$ (erg s$^{-1}$ ), or log SFR = log $L([CII])$ - 6.93 $\pm$ 0.15 for SFR in \mdot~ and $L([CII])$ in \ldot.  Our result derived in Paper 1 from calibrating SFR using the total infrared dust luminosity was log SFR = log $L([CII)])$ - 7.08 $\pm$ 0.3, so the two independent results agree well within the cosmic variance among sources.  The systematic difference in SFR calibration is less than the variance within either calibration.  From this, we conclude that the relation log SFR = log $L([CII)])$ - 7.0 can measure SFR to a precision of $\pm$ $\la$ 50\% for any individual source within the uncertainties arising from cosmic variance.

Our conclusion that [CII] is a reliable SFR indicator differs from the conclusions of \citet{dia13} and \citet{far13}.  This difference arises primarily from our calibration of SFR derived from either the bolometric luminosities or the Neon line luminosities using only pure starbursts based on the PAH classification criterion. The spectral energy distributions of dusty sources show large variations, with changes in characteristic temperatures that may depend on luminosity or redshift \citep[e.g.][]{lof13,mel12}.  Determining the relative AGN/starburst contribution to dust luminosity at different infrared wavelengths is a complex problem to which various diagnostics can be applied \citep{vei09}.  By using only a subset of sources having no indications of any AGN contribution, contamination from AGN luminosity cannot affect the calibration of SFR, and it is not necessary to determine quantitatively the AGN/starburst fraction within individual sources to determine a SFR calibration.  This issue and the differences from previous conclusions are considered further in the next section.

All of the results summarized above show consistent correlations between the [CII] luminosity and the luminosity of the starburst PDRs measured using the [CII]/PAH ratio and from the starburst HII regions measured with the Neon lines. These results imply that [CII] measures the same starbursts as are measured with the mid-infrared diagnostics.  Are there any sources with indications of excess [CII] luminosity that may not be associated with the starburst PDR or HII region luminosities?  Additional [CII] could arise from lower ionization regions associated with an AGN or from widespread low ionization from older starbursts and association with infrared cirrus \citep{lof13}.  There could also be excess [CII] compared to the mid-infrared diagnostics if starbursts are so dusty that the mid-infrared features suffer significant extinction compared to [CII] (e.g. Farrah et al. 2007, 2013).

The ratios among [CII]/PAH/[NeII] in Figures 3 and 4 show median ratios that are the same for all classifications from AGN through starburst. The sources with the largest [CII] excess are among composites and AGN, so those are the best candidates for sources with additional [CII] not associated with the starbursts. For this reason, we attempt an estimate of the excess [CII] luminosity which may be present in composites and AGN.  

To do this, we determine the [CII] luminosity remaining in the AGN and composite sources, $L([CII],other)$, after subtracting from the total $L([CII])$ the luminosity component from starbursts, $L([CII],SB)$, assuming the median observed ratio log $L([CII],SB)$/$L(PAH~11.3)$ = 0.22 for starbursts in Figure 3. For example, if log $L([CII],other)$/$L([CII])$ = -0.3, this means that 50\% of the observed $L([CII])$ does not arise from the starburst component.  In many cases, this leads to $L([CII],SB)$ $>$ $L([CII])$.  In these cases, we arbitrarily adopt a limit $L([CII],other)$ $<$ 0.1~$L([CII])$. 

Because of the scatter of $\sim$ 0.2 dex in the $L([CII],SB)$/$L(PAH~11.3)$ ratio for individual starbursts (Figure 3), only statistical estimates for the overall samples can be determined in this way. Of the 36 composite sources, only 4 have log $L([CII],other)$/$L([CII])$ $>$ -0.3.  This means that [CII] luminosity from alternative sources may exceed [CII] luminosity from conventional starbursts in only 11\% of composite sources.  For AGN, only 14 of 60 AGN exceed log $L([CII],other)$/$L([CII])$ = -0.3, meaning that 23\% of AGN may have excess [CII] luminosity that exceeds the starburst [CII] luminosity.   While these estimates are approximate and qualitative, they indicate that there is no strong evidence within our sample for sources in which [CII] luminosity arising from other mechanisms exceeds the [CII] luminosity from ongoing star formation

%\figurenum{13}
%\includegraphics[scale=0.9]{figure13May2013.ps}
%\caption{(new-figure2.ps)Equivalent width of [CII] emission line, with uncertainties, compared to far infrared continuum luminosity vLv(157).  Uncertainties are shown by plotting the one sigma upper and lower values of the continuum flux, using same fractional uncertainty as for EW.  Reason is because cannot reliably sum continuum over 3x3 pixels, so determine continuum from the EW of the line, which is summed over 3x3 and the EW uncertainty is a measure of the continuum uncertainty.}  
%\end{figure}

\subsection {Ratio of [CII] Luminosity to Continuum Luminosities}

%NOTE ELBAZ 2011 FOR SUMMARY OF SFR  QUESTION.   AND USE OF FAR IR.  PAGE 2012 (NATURE) FOR FIR TO SFR.  SO IMPORTANT TO PURSUE BELOW QUESTIONS.  SEE ZHAO 2013 AND DIAZ-SANTOS 2013 FOR CII/LIR, AND WHETHER HAVE COMPACT, DUSTY STARBURSTS.  LOW RATIOS HAVE HIGH IONIZATION PARAMETERS.  SIMPSON 2013 HIGH CII/FIR AT Z ~ 4.   DAI 2012 FOR MINORITY OF QSOS WHICH DO HAVE FIR (IN LOCKMAN) Coppin? Alexander??

The comparisons of [CII] with Neon line fluxes in the preceding section confirm, as did the PAH comparisons discussed in Paper 1, that the [CII] luminosity gives the same measure of SFR as is found using the mid-infrared diagnostics.  A related and important question is how SFR measured with [CII] compares with SFR measured using only the infrared continuum luminosity.  This is crucial because of the common use of the continuum luminosity alone, particularly the far infrared continuum, to measure the evolution of star formation in the universe \citep{elb11,zha13}.  

The biggest question in the comparison of continuum luminosity and $L([CII)])$ arises from previous observations that more luminous sources in $L_{fir}$ have relatively weaker $L([CII)])$, termed the ``[CII] Deficit" \citep{luh03,hel01,gra11,sta10}.  The $L_{fir}$ is usually defined as the luminosity within 40 \um $\la$ $\lambda$ $\la$ 125 \ums, measured quantitatively using the 60 \um and 100 \um photometry with IRAS according to the formulation in \citet{hel88}. A deficit could be explained by starbursts with increasing ionization parameter and harder ionizing radiation in more luminous sources \citep{mal01,abe09,sta10} or by having dustier, more obscured starbursts \citep{far13,dia13}.  If $L_{fir}$ invariably measures SFR, then a deficit would mean that the SFR from [CII] is systematically underestimated for the most luminous sources.  

In Paper 1, we concluded that the deficit is not a consequence of differences among starbursts but instead arises because more luminous sources have increasing contributions from AGN to the dust continuum luminosity. This conclusion arose because luminous sources with an apparent deficit show the mid-infrared diagnostic of AGN (weak PAH), indicating that most continuum luminosity is produced by AGN dust heating.  This is illustrated in Figure 9 of Paper 1, where our sources reach $L_{ir}$ $\ga$ 10$^{13}$ \ldot, but all sources except one having $L_{ir}$ $>$ 10$^{12}$ \ldot~ contain AGN.  The one starburst does not show a deficit compared to lower luminosity starbursts, but the high luminosity AGN do show a systematic deficit. 

We illustrate this result again in Figure 7 using the conventional parameters for the deficit, the ratio $L([CII)])$/$L_{fir}$ compared to $L_{fir}$, determining $L_{fir}$ from the IRAS fluxes tabulated in \citet{sar11}.  The plot is made only for pure starbursts, as defined by our classification criterion EW(PAH 6.2 \ums) $>$ 0.4 \ums.  The distribution of results for our sample (crosses) confirms the conclusion of Paper 1 - that no trend of ratio with luminosity is seen in the $L([CII)])$/$L_{fir}$ ratio when only pure starbursts are used.  

Figure 7 also includes results (circles) for starbursts in the [CII] sample of \citet{dia13} defined by their stated criterion for pure starbursts, which is EW(PAH 6.2 \ums) $>$ 0.5 \um using PAH measurements from \citet{sti13}.  These results show an overall median that agrees with ours.  Within the limited range of $L_{fir}$, any trend for $L([CII)])$/$L_{fir}$ to decrease with $L_{fir}$ in the Diaz-Santos sample is less than the one sigma dispersion among starbursts at any given luminosity.  For example, from log $L_{fir}$(\ldot) = 10.5 to log $L_{fir}$(\ldot) = 11.5, the median ratio decreases by a factor of two, but the dispersions at both luminosities exceed a factor of 2.5.  Their sources with 10 $<$ log $L_{fir}$(\ldot) $<$ 11 exceed our median ratio by only a factor of 1.5, and their sources with 11 $<$ log $L_{fir}$(\ldot) $<$ 12 are below our median ratio by a factor of 0.8.  These comparisons indicate consistency between the two samples of starbursts.  

%OMIT? Nevertheless, it remains possible to postulate the existence of "hidden" starbursts having weak [CII] that account for most $L_{fir}$ if these starbursts mimic AGN in the mid-infrared by having weak PAH emission \citep{dia13}. If such starbursts exist with no detectable [CII], [NeII], or PAH spectral signature, these starbursts should nevertheless produce excess far infrared luminosity from cool dust compared to the luminosity associated with starbursts which show the signatures.  To minimize contamination by AGN, which we have demonstrated can affect $L_{fir}$, a test for such excess far infrared luminosity   

To consider further the deficit issue, we can extend consideration of the [CII] line to continuum ratio to longer wavelengths (representing emission from cooler dust) than measured by $L_{fir}$.  This can be done because the continuum at rest frame 158 \um is also measurable for most of our sources having [CII] line observations.  Attributing a deficit to luminous starbursts invokes the assumption that the coolest dust in luminous sources is invariably heated by starbursts instead of AGN.  Even if the far-infrared continuum at $\la$ 100 \um can include contributions from AGN heating, as we have concluded, it might be expected that cooler dust seen at longer wavelengths would be increasingly dominated by the starburst component.  If a deficit really exists among pure starbursts, therefore, the deficit should be as readily seen comparing $L([CII)])$ to $\nu L_{\nu}$(158 \ums) as when comparing to $L_{fir}$.  Conversely, if the apparent deficit arises because $L_{fir}$ is contaminated by hotter AGN dust, any deficit arising because AGN are included in the sample should diminish when measured with the cooler dust seen at 158 \ums.

\begin{figure}
\figurenum{7}
\includegraphics[scale=0.9]{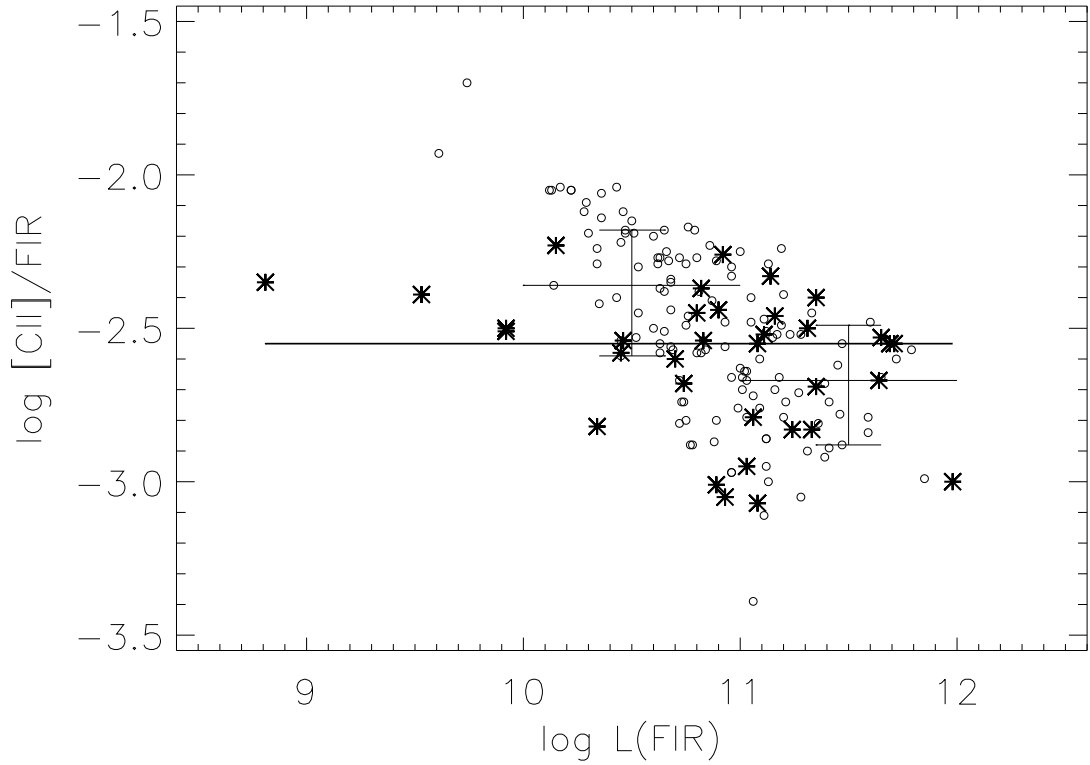}
\caption{Ratio of L([CII])/$L_{fir}$ compared to $L_{fir}$ for starbursts in \ldot.  Asterisks are starbursts from current sample and thick horizontal line is their median.  Circles are starbursts from \citet{dia13} using their criterion for pure starbursts; thin horizontal lines and error bars are medians and one sigma dispersions within this sample for 10 $<$ log $L_{fir}$ $<$ 11, and 11 $<$ log $L_{fir}$ $<$ 12.}
\end{figure}

\begin{figure}
\figurenum{8}
\includegraphics[scale=0.9]{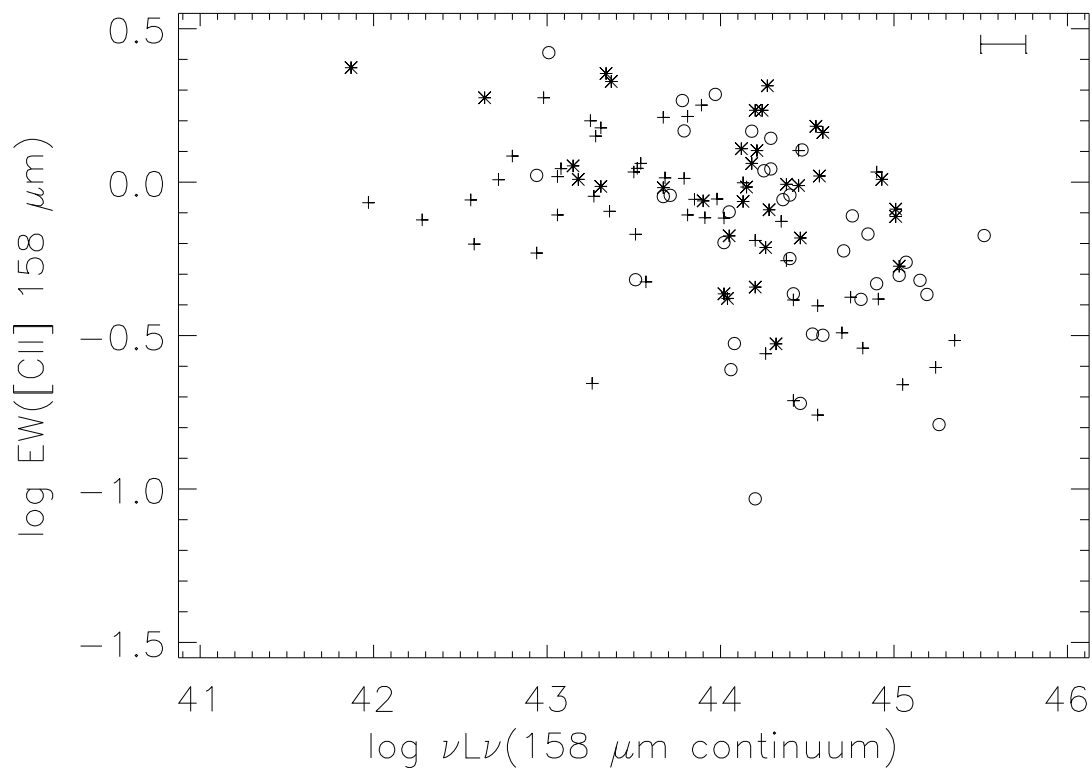}
\caption{Far infrared continuum luminosity at 158 \ums, log $\nu L_{\nu}$(158 \ums) in erg s$^{-1}$, compared to equivalent width of [CII] emission line in \um for all sources in Table 2. Crosses are AGN from the EW(PAH 6.2 \ums) classification, open circles are composite AGN plus starburst, and asterisks are starbursts.  Error bar shows median uncertainty in luminosity determined by observational uncertainty of continuum flux measurement. }  
\end{figure}

%\begin{figure}

The strength of [CII] compared to the continuum is defined by the equivalent width (EW), which is a linear measurement of the line to continuum ratio.  EW relates to continuum luminosities by $\nu L_{\nu}$(158 \ums) = $\lambda L_{\lambda}$(158 \ums) = $\lambda$$L([CII])$/ EW([CII]) for $\lambda$ = 158 \ums.  Measuring $L([CII)])$ and $\nu L_{\nu}$(158 \ums) with the same PACS observation has an additional advantage in that the spatial resolution of both measures is identical.  For sources that may be resolved, this removes one concern when comparing $L([CII)])$ with $L_{fir}$ determined from IRAS photometry because the IRAS spatial resolution is much poorer than the spaxels of PACS. 

The [CII] line fluxes presented in section 2.2 are measured by summing over 9 spaxels.  The continuum flux densities are much weaker than the line and are often so weak in outlying spaxels that they are overwhelmed by noise.  To minimize the effects of continuum noise, the continuum $f_{\nu}$(158 \ums) is determined by using the total $f$([CII)]) from Table 2 combined with EW([CII]) measured only in the brightest spaxel.  These EW are measured in all of our spectra and given in Table 2.  The uncertainties are often large, primarily because the continua are faint and have large uncertainties after backgrounds are subtracted.  The EW uncertainties that are given in Table 2 include the combined uncertainty of the total line flux and of the continuum level in the brightest spaxel, but uncertainties are dominated by the uncertainty in the continuum.  

The $\nu L_{\nu}$(158 \ums) is compared to EW([CII]) in Figure 8 for all of the [CII] detections in Table 2 including AGN, composites and starbursts (the 10 non-detections are not used for this plot because their limits apply to both coordinates).  Coordinates are in log units to make direct comparisons with the line to continuum ratio plots $L([CII)])$/$L_{fir}$ used to discuss the deficit in previous references.  Figure 8 shows similar ranges and dispersions of both values in all of the distributions for AGN, composites, and starbursts.  The cosmic variance among individual sources is greater than any systematic differences between classes (shown more quantitatively in the discussion below of Figure 9). 

The overall conclusion from Figure 8 is that the appearance of a ``deficit" is much less conspicuous when using $\nu L_{\nu}$(158 \ums) than when using $L_{fir}$.  For example, comparisons can be made with Figure 1 of \citet{dia13} that compares $L([CII)])$/$L_{fir}$ to $L_{fir}$ for their combined sample of AGN, composites, and starbursts.  The luminosity range is comparable to that in Figure 8, $\ga$ 10$^{3}$, but the range of line to continuum ratio is much greater in the comparison with $L_{fir}$ ($\sim$ 100), compared to a range of $\sim$ 10 in Figure 8. Another conspicuous difference is the clear trend for the upper envelope of points to show a deficit when $L([CII])$ is compared to $L_{fir}$, but this trend is not apparent in Figure 8.  Some high luminosity sources show EW as large as some low luminosity sources.  These results indicate that the apparent deficit becomes less significant when continuum luminosity is measured at the longest wavelengths, consistent with the interpretation that the coolest dust arises in the same starbursts measured with the [CII], PAH, and Neon features for starbursts of all luminosities. 

This conclusion can be further illustrated by comparing EW([CII]) with source classification based on EW(PAH 6.2 \ums); if $\nu L_{\nu}$(158 \ums) and L([CII)]) both measure only the starburst component of any source, the EW([CII]) should be independent of source classification.  The results are shown in Figure 9, showing all EW([CII]) in Table 2. The scatter increases for AGN compared to starbursts primarily because of increased uncertainties caused by fainter continua.  For starbursts, the typical observational uncertainty in EW(PAH 6.2 \ums) given in Table 2 is only $\sim$ 10\% because the lines and continua are strong, but this uncertainty rises to $\sim$ 50\% for the weak lines and continua of the AGN.

%[[below moved up from next section - point  out that these sections have been rerranged to put the deficit and EW discussion in the same section]]

\begin{figure}
\figurenum{9}
\includegraphics[scale=0.9]{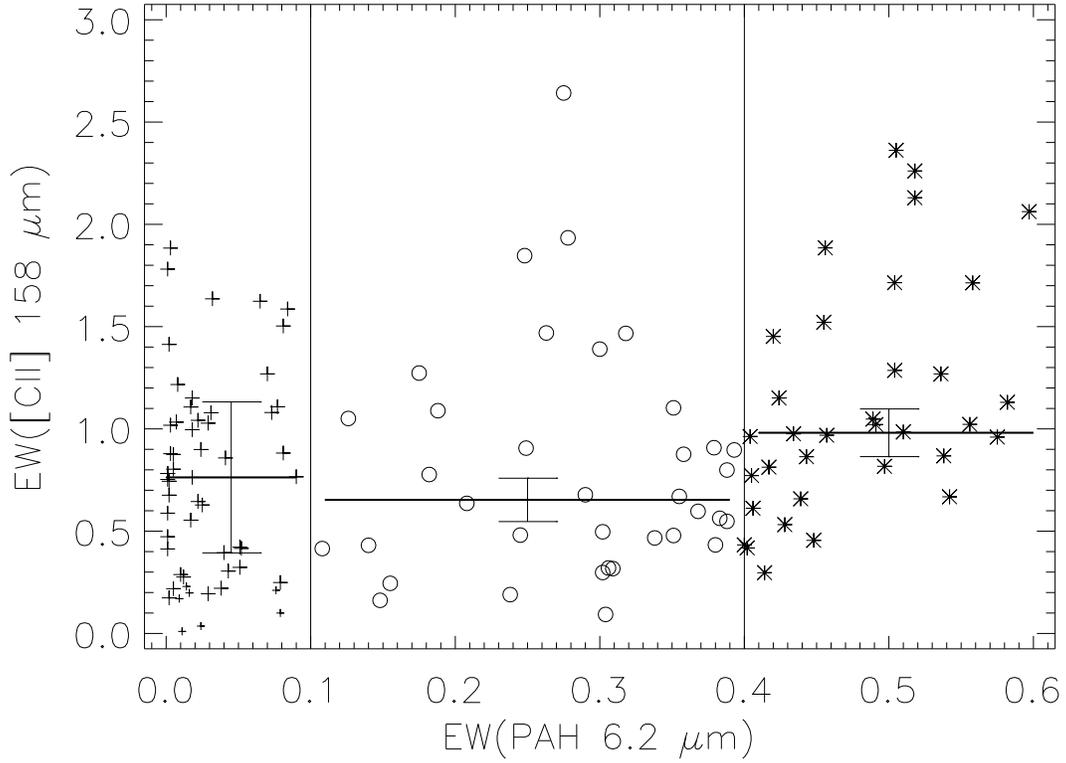}
\caption{Equivalent width of [CII] 158 \um emission line in \ums, with uncertainties, compared to EW(PAH 6.2 \ums) classification, in \ums.  Medians are shown as horizontal lines in each class.  Error bars show median one sigma uncertainties of the individual EW measurements for different classifications using uncertainties for each source given in Table 2.  Sources with upper limits in both EWs are not plotted or included in medians, but sources with upper limits only in EW([CII]) are shown as small symbols (all AGN). }  
\end{figure}

The crucial result in Figure 9 is that the median EW([CII]) is only marginally greater for starbursts compared to composites and AGN; within the uncertainty, a median value of EW([CII]) = 0.9 \um falls within all error bars. For starbursts, the median  EW([CII]) = 1.0 \ums, decreasing to EW([CII]) = 0.7$\pm$0.4 \um for AGN, and the difference is smaller than the dispersion in EW([CII]) among individual starbursts.  The EW([CII]) changes much less than EW(PAH 6.2 \ums).  The EW(PAH) ranges from $<$ 0.01 \um to 0.6 \ums, a range of more than 60.  This happens because an AGN has greatly increased mid-infrared continuum from warm dust compared to a starburst, thereby decreasing EW(PAH 6.2 \ums).  By contrast, the EW([CII]) ranges from $\sim$ 0.2 \um to $<$ 2 \ums, a range of $\sim$ 10.  This result indicates quantitatively that the far infrared continuum at 158 \um scales much more closely with the starburst component of any source than does the continuum at shorter wavelengths. This result illustrates why the $\nu L_{\nu}$(158 \ums) could be a better estimate of SFR regardless of source classification than is $L_{fir}$, which includes shorter wavelength emission from hotter dust arising around AGN. 

\subsection{SFR from the 158 \um Continuum}

%[[rearrange and moved up to this section.  Respond that this topic is now given its own section]] 

The preceding section shows that there is no conspicuous deficit when $L([CII)])$ is compared to $\nu L_{\nu}$(158 \ums) for the full sample of AGN, composites and starbursts over all luminosities.  Any systematic change in $L([CII)])$/$\nu L_{\nu}$(158 \ums) with luminosity or with classification is small compared to the the cosmic variance at a given luminosity.  This result encourages the conclusion that the far infrared luminosity at 158 \um scales similarly with $L([CII)])$ in sources of all classifications and luminosity.  This would make $\nu L_{\nu}$(158 \ums) a more reliable measure of SFR than the $L_{fir}$, if the latter sometimes includes an AGN component of luminosity.  
 
It was concluded in section 3.2  that log SFR = log $L([CII])$ - 6.93 $\pm$ 0.15 for SFR in \mdot~ and $L([CII])$ in \ldot, calibrated by comparison with the [NeII] and [NeIII] luminosities.  The result derived independently in Paper 1 by comparing [CII] luminosities with SFR derived from total infrared dust luminosity as in \citet{ken98} was log SFR = log $L([CII)])$ - 7.08 $\pm$ 0.3.   Both results agree within the uncertainties, so we take as an overall calibration combining these results that log SFR = log $L([CII)])$ - 7.0$\pm$ 0.2.  In both cases, the quoted uncertainties show only the cosmic variance among different starbursts but do not reflect any systematic uncertainties in the underlying calibrations. The agreement of the two independent methods implies, however, that such systematic calibration uncertainties are smaller than the variance among sources.  

Results for the EW([CII]) in Figure 9 allow a transformation from SFR as measured from $L([CII)])$ to SFR measured from $\nu L_{\nu}$(158 \ums). This leads to a comparison with another independent calibration.  Using the median EW([CII]) of 1.0 \um for starbursts alone to mimimize any residual contamination by AGN in the continuum, the relations among EW, $L([CII)])$, SFR, and $\nu L_{\nu}$(158 \ums) yield log SFR = log $\nu L_{\nu}$(158 \ums) - 42.8$\pm$ 0.2 for SFR in \mdot~ and $\nu L_{\nu}$(158 \ums) in erg s$^{-1}$ (to match units in Figure 8).  For the luminosities of our sample shown in Figure 8, the resulting SFRs range over 0.8 $<$ log SFR $<$ 2.7.   

%$\nu L_{\nu}$(158 \ums) = $\lambda$$L([CII])$/ EW([CII]),      log $\nu L_{\nu}$(158 \ums) = 2.20 +logL(CII) = 2.20 + logSFR + 7.0 = 9.2 + log SFR or log SFR = log $\nu L_{\nu}$(158 \ums) -9.2 for lum in solar, but becomes log SFR = log $\nu L_{\nu}$(158 \ums) -42.8         for lum in erg per sec, as in Figure 9 

A previous effort with the same objective but a completely independent calibration was by \citet{cal10}, who used 160 \um photometry with $Spitzer$ for comparing continuum $\nu L_{\nu}$(160 \ums) to the SFR calibrated by H$\alpha$ luminosity.  Their sample consisted of 189 nearby star forming galaxies with generally low infrared luminosities which are not obscured LIRGs or ULIRGs so that SFRs could be calibrated using H$\alpha$ luminosities without any corrections for extinction. They caution that lower luminosity systems may have significant 160 \um luminosity from dust heated by stellar populations not associated with the current star formation, but they conclude that the $\nu L_{\nu}$(160 \ums) is dominated by ongoing star formation for $\nu L_{\nu}$(160 \ums) $>$ 2 x 10$^{42}$ erg s$^{-1}$ (which encompasses all of our sources).  Their resulting calibration is log SFR = log $\nu L_{\nu}$(160 \ums) - 42.85.  This is nearly identical to our result even though derived with completely independent parameters.  

The agreement is somewhat fortuitous because Calzetti et al. use different relations between ionizing photons and SFR compared to those of Kennicutt (1998) which were used for our calibration of SFR using bolometric luminosities.  Calzetti et al. note that with the Kennicutt calibrations, their result would be log SFR = log $\nu L_{\nu}$(160 \ums) - 42.68.  Although this gives a SFR measure from the 160 \um continuum that is formally 1.3 times larger than our result, it is within our uncertainties.  The combined results of the two samples cover a factor of 100 in SFRs and a wide range of dust obscuration, implying a generalized validity to use of the 160 \um continuum as a measure of SFR so long as there is confidence that the dust continuum is dominated by stellar heating. 

A calibration of SFR from $L_{\nu}$(158 \ums) was also determined by \citet{far13} for a sample of 25 ULIRGs (dominated by AGN) for which they observe [CII] and continuum fluxes. Their calibration of SFR arises from an earlier comparison \citep{far07} between PAH(6.2 \um + 11.3 \ums) luminosities and SFRs determined from [NeII] and [NeIII] luminosities using the results of \citet{ho07}.  By comparing SFRs determined from these PAH luminosities with the observed $L_{\nu}$(158 \ums) for the 25 ULIRGs, they determine an empirical relation between SFR and continuum luminosity density that log SFR = 3.36$\pm$0.22 + (1.42$\pm$0.30)log $L_{\nu}$(158 \ums), for luminosity density in \ldot~ per Hz.  The non linear luminosity dependence implies that a greater SFR is measured from continuum luminosity compared to PAH luminosity as luminosity increases, which is consistent with having a residual contamination of $L_{\nu}$(158 \ums) from AGN luminosity for increasing luminosities.  Within the uncertainties of this fit and the luminosity dependence, our results for SFRs are similar.  For example, at log $\nu L_{\nu}$(158 \ums) = 45 in erg s$^{-1}$, or log $L_{\nu}$(158 \ums) = -0.86 in \ldot~per Hz, our calibration from pure starbursts yields log SFR = 2.2$\pm$ 0.2 compared to the Farrah et al. result of 1.7 $<$ log SFR $<$ 2.6.  

The uncertainty within our results that is introduced when using $\nu L_{\nu}$(158 \ums) to measure SFR compared to using $L([CII)])$ in sources of different classification can be estimated from the differences among EW([CII]) for different classifications in Figure 9.  If some of the $\nu L_{\nu}$(158 \ums) for AGN and composites arises from dust heating by an AGN, then EW([CII]) should be less for these sources compared to starbursts.  Adopting the median EW([CII]) = 1.0 \um as the measure of line to continuum ratio for pure starbursts leads to an estimate of the SFR error that would derive from use only of the continuum for AGN or composite sources.  The error in SFR measure is inversely proportional to the EW; for example, an EW([CII]) = 0.5 \um would mean a continuum twice as strong as from starbursts, so the continuum alone would lead to an overestimate of SFR by a factor of two compared to the SFR measured from [CII]. For AGN, the median EW([CII]) = 0.7 \ums, which implies a systematic overestimate of SFR by a factor of 1.4 if using only the far infrared continuum as the SFR indicator in AGN.   For AGN, 21/56 have EW([CII]) $<$ 0.5 \ums, leading to an overestimate of SFR from the continuum alone by more than a factor of two in $\sim$ 40\%.  Although these results lead to smaller discrepancies in SFR than would be measured using only $L_{fir}$, confidence in measuring SFR within a factor of two from continuum luminosities alone requires confidence that the source is dominated by a starburst without continuum contamination by an AGN.

\section {Summary and Conclusions}

New [CII] 158 \um observations of 18 sources with the $Herschel$ PACS instrument are presented, and a summary of our total sample of 130 [CII] sources is given that covers a wide range of AGN to starburst classifications as derived from PAH strength.  New redshifts derived from [CII] and line to continuum strengths (equivalent width of [CII]) are given for the full sample.  

Results for 112 sources are compared with emission line fluxes from high resolution $Spitzer$ IRS spectra.  A new calibration of [CII] as a SFR indicator is determined by comparing [CII] fluxes with mid-infrared [NeII] and [NeIII] emission line fluxes.  This independently gives the same result as determining SFR using bolometric luminosities of reradiating dust from starbursts:  log SFR = log $L([CII)])$ - 7.0$\pm$0.2, for SFR in \mdot~ and $L([CII])$ in \ldot.  This confirms that [CII] measures the same starburst component of sources as measured with mid-infrared PAH and Neon emission line diagnostics.

The line to continuum ratio measured at 158 \ums, EW([CII]), changes little with luminosity or with classification, indicating that the far infrared continuum at 158 \um arises primarily from the starburst component of any source.  For pure starbursts, the continuum alone gives log SFR = log $\nu L_{\nu}$(158 \ums) - 42.8$\pm$0.2 for SFR in \mdot~ and $\nu L_{\nu}$(158 \ums) in erg s$^{-1}$. The change of EW([CII]) with classification (median EW([CII]) = 1.0 \um for starbursts compared to 0.7 \um for AGN) implies a systematic overestimate of SFR in AGN by a median factor of 1.4 if using only the far infrared continuum at 158 \um as a SFR indicator.

\acknowledgments

We thank those who developed the $Herschel$ Space Observatory for the opportunity to observe with open time.  PACS has been developed by a consortium of institutes led by MPE (Germany) and including UVIE (Austria); KU Leuven, CSL, IMEC (Belgium); CEA, LAM (France); MPIA (Germany), INAF-IFSI/OAA/OAP/OAT, LENS, SISSA (Italy); and IAC (Spain). This development was supported by the funding agencies BMVIT (Austria), ESA-PRODEX (Belgium), CEA/CNES (France), DLR (Germany), ASI/INAF (Italy), and CICYT/MCYT (Spain). Partial support for this work at Cornell University was provided by NASA through Contracts issued by the NASA $Herschel$ Science Center. We thank Dieter Engels and the Hamburger Sternwarte for hospitality during the initial phase of this effort.

\end{document}